\documentclass[transaction]{IEEEtran}
\usepackage{graphicx}
\usepackage{epstopdf}
\usepackage[latin1]{inputenc}
\usepackage{amsmath,amssymb,amscd,latexsym,dsfont}
\usepackage[english]{babel}
\usepackage{tabularx,cite}
\usepackage{graphicx}
\usepackage{algpseudocode}
\usepackage{algorithm}
\usepackage{algorithmicx}
\usepackage{float}
\usepackage{multicol}
\usepackage{psfrag}
\usepackage{comment,psfrag,subfigure,enumerate}
\usepackage{color}
\usepackage{amsthm}
\usepackage{graphicx}
\usepackage{graphicx}
\usepackage{epstopdf}
\usepackage[latin1]{inputenc}
\usepackage{amsmath,amssymb,amscd,latexsym,dsfont}
\usepackage[english]{babel}
\usepackage{tabularx,cite}
\usepackage{graphicx}
\usepackage{algpseudocode}
\usepackage{algorithm}
\usepackage{algorithmicx}
\usepackage{float}
\usepackage{multicol}
\usepackage{mathtools}
\usepackage{psfrag}
\usepackage{comment,psfrag,subfigure,enumerate}
\usepackage{color}
\usepackage{amsthm}
\usepackage{multirow}

\ifCLASSINFOpdf
\else
\fi
\begin{document}
\IEEEoverridecommandlockouts
\IEEEpubid{\makebox[\columnwidth]{978-1-4799-5863-4/14/\$31.00 \copyright 2014 IEEE \hfill} \hspace{\columnsep}\makebox[\columnwidth]{ }}
\title{On the Throughput of Large-but-Finite MIMO Networks using Schedulers}

%\author{\IEEEauthorblockN{Authors}\\
%\IEEEauthorblockA{Department of Signals and Systems,
%Chalmers University of Technology, Gothenburg, Sweden\\
%Email: \{behrooz.makki, alexandre.graell, thomase\}@chalmers.se}
%%\and
%%\IEEEauthorblockN{Thomas Eriksson}
%%\IEEEauthorblockA{Department of signals and systems\\
%%Chalmers University of Technology\\
%%Gothenburg, Sweden\\
%%Email: thomase@chalmers.se}
%\thanks{Alexandre Graell i Amat was supported by the Swedish Agency for Innovation Systems (VINNOVA) under the P36604-1 MAGIC project.}
%\thanks{Behrooz Makki and Tommy Svensson are with Chalmers University of Technology, Email: \{behrooz.makki, tommy.svensson, thomase\}@chalmers.se. Mohamed-Slim Alouini is with the King Abdullah University of Science and Technology (KAUST), Email: slim.alouini@kaust.edu.sa}
%\thanks{Part of this work has been submitted for possible presentation at the IEEE ICC 2016.}
%}
\author{\IEEEauthorblockN{Behrooz Makki, Tommy Svensson, and Mohamed-Slim Alouini, \emph{Fellow, IEEE}}\\
%\IEEEauthorblockA{Department of Signals and Systems,
%Chalmers University of Technology, Gothenburg, Sweden\\
%Email: \{behrooz.makki, alexandre.graell, thomase\}@chalmers.se}
%%\and
%%\IEEEauthorblockN{Thomas Eriksson}
%%\IEEEauthorblockA{Department of signals and systems\\
%%Chalmers University of Technology\\
%%Gothenburg, Sweden\\
%%Email: thomase@chalmers.se}
%\thanks{Alexandre Graell i Amat was supported by the Swedish Agency for Innovation Systems (VINNOVA) under the P36604-1 MAGIC project.}
%\thanks{Author 1, Author 2 and Author 3 are with ... Author 4 is with ...}
\thanks{Behrooz Makki and Tommy Svensson are with Chalmers University of Technology, Gothenburg, Sweden, Email: \{behrooz.makki, tommy.svensson\}@chalmers.se. Mohamed-Slim Alouini is with the King Abdullah University of Science and Technology, Thuwal, Saudi Arabia, Email: slim.alouini@kaust.edu.sa}
%\thanks{The research leading to these results received funding from the European Commission H2020 programme under grant agreement $n^{\circ}$671650 (5G PPP mmMAGIC project), and from the Swedish Governmental Agency for Innovation Systems (VINNOVA) within the VINN Excellence Center Chase.}
\thanks{Part of this work has been {presented at the IEEE WCNC 2018.}}
}
%Telephone: (800) 555--1212, Fax: (888) 555--1212}
\maketitle
%\onecolumn
\vspace{-10mm}
\begin{abstract}
This paper studies the sum throughput of the {multi-user} multiple-input-single-output (MISO)  networks in the cases with large but finite number of transmit antennas and users. Considering continuous and bursty communication scenarios with different users' data request probabilities, we derive quasi-closed-form expressions for the maximum achievable throughput of the networks using optimal schedulers. The results are obtained in various cases with different levels of interference cancellation. Also, we develop an efficient scheduling scheme using genetic algorithms (GAs), and evaluate the effect of different parameters, such as channel/precoding models, number of antennas/users, scheduling costs and power amplifiers' efficiency, on the system performance. Finally, we use the recent results on the achievable rates of finite block-length codes to analyze the system performance in the cases with short packets. As demonstrated, the proposed GA-based scheduler reaches (almost) the same throughput as in the exhaustive search-based optimal scheduler, with substantially less implementation complexity. {Moreover, the power amplifiers' inefficiency and the scheduling delay affect the performance of the scheduling-based systems significantly}.
\end{abstract}
%Then, compared to open-loop communication setups, the implementation of power-adaptive ARQ reduces the average power by ? and ? dB, if a maximum of 2 and 3 retransmissions is utilized, respectively.
% \fi
\IEEEpeerreviewmaketitle
\vspace{-1mm}
\section{Introduction}
The next generation of wireless communication networks must provide data streams for everyone everywhere at any time.
%To address these demands, the main strategy persuaded in the last few years is network \emph{densification} \cite{7801976,m06094252,m07127500,m06884146,m06831724,06095627,m07037316,m06941317}.
One of the promising techniques to {address these demands} is to use many {antennas} at the transmitter and/or receiver sides. This approach is referred to as massive or large multiple-input-multiple-output (MIMO) in the literature.

In general,  the network sum rate increases with the number of transmit/receive {antennas}. Thus, the trend is towards  { base stations and/or users with} asymptotically {large number of antennas}. However, large MIMO implies challenges such as hardware impairments and signal processing complexity which may limit the number of antennas in practice. Therefore, it is interesting to analyze the performance of the multi-user MIMO systems in the cases with large but finite number of transmit {antennas} and/or users. Particularly, {user} scheduling algorithms, in which only a set of users are activated based on the channel quality/metric of interest, are appropriate approaches to utilize the diversity of large MIMO systems and optimize the network performance.

The scheduler-based data transmission of multi-user MIMO systems is studied in, e.g., \cite{8377145,1705935,5360758,1492685,04668538,m06094252,m07127500,m06884146,m06831724,06095627,m07037316,m06941317,06850064,05456039,05336871,01312486,01664083,04635028,04570215,04299613,05374079,05871799,06504540,06878495,06740115,06725595,05999732,06810624,04784356}. {In \cite{8377145}, we study the performance of a genetic algorithm (GA)-based scheduler in the cases with large-but-finite number of transmit antennas/users. Then,} {\cite{1705935,5360758,1492685,04668538} develop GA-based schedulers for MIMO systems in the cases with zero-forcing beamforming \cite{1705935,5360758,1492685} and dirty paper coding \cite{04668538}. Also,} \cite{m06094252,m07127500,m06884146,m06831724,06095627,m07037316,m06941317} study the problem in massive MIMO systems with asymptotically large number of {transmit antennas/receivers}. The scheduling algorithms are mostly designed to maximize the network sum rate/capacity \cite{8377145,5360758,1492685,04668538,06850064,05456039,05336871,01312486,01664083,04635028,04570215,04299613,05374079,05871799,06504540,06095627,06878495,06740115,m06094252,m06884146,m07037316,m06941317,m06831724,m07127500}, while other objective functions such as {the coverage area \cite{1705935},} the feedback load \cite{01312486}, the symbol error probability \cite{01664083}, the packet drop rate \cite{06878495}, the outage probability \cite{06725595}, the bit error probability \cite{06725595},  the outage capacity \cite{05999732} and the spectral efficiency \cite{06810624}   have been considered as the objective function as well. Moreover, e.g., \cite{1492685,04784356,06095627,m06094252,04668538} suggest different algorithms for fair scheduling. Finally, the effect of imperfect channel state information (CSI) on the performance of the scheduling techniques has been investigated by \cite{m06831724,m07037316,m06941317,06850064,06725595,06095627,05456039,05336871,04784356,01312486}, and different limited feedback methods have been proposed.

As highlighted in the literature, depending on the power constraints, precoding schemes and the considered metrics, user scheduling may be a non-convex problem. Also, unless {for the cases with few transmit antennas and/or users,} the search space for the scheduler optimization is extremely large which prohibits an optimal search approach, and even makes the simulations challenging. Example 1 demonstrates some typical numbers for the scheduling complexity as a motivation for our problem formulation and analysis.

\textbf{\emph{Example 1: On the Scheduling Complexity.}} To motivate for deriving an efficient scheduling algorithm and closed-form expressions for the maximum achievable throughput of the scheduler-based large-but-finite MIMO networks, consider a setup with $M=30$ transmit antennas and a maximum of $\tilde N=100$ users each with data requesting probability $\alpha=0.8.$ Let ${n \choose k}$ denote the ``\emph{n choose k}'' operator.
 %and identical long-term channel statistics of the users.
 If in a time slot $N=80$ users ask for data transmission, with probability ${\tilde N \choose N}\alpha^N(1-\alpha)^{\tilde N-N}=0.099$, the scheduler needs to check ${80 \choose 30}\simeq 8.8 \times 10^{21}$ possible user scheduling {selection schemes} {in that specific time slot}. Also, the expected number {of checkings}, averaged over multiple times slots, is
\vspace{-0mm}
\begin{align}\label{eq:eqexpectedcheckings}
\bar S=\sum_{i=1}^M{{\tilde N \choose i}\alpha^i(1-\alpha)^{\tilde N-i}}+\sum_{i=M+1}^{\tilde N}{\tilde N \choose i}\alpha^i(1-\alpha)^{\tilde N-i}{i \choose M},
\end{align}
which with the considered parameter settings leads to $\bar S\simeq 3.6\times10^{22}$. {Then,} to derive the expected throughput, we need to run the scheduling algorithm for, say, $ 10^{6}$ different time slots, i.e., in total $\simeq 3.6\times 10^{28}$ {checkings}, which is not feasible in a limited time.

In this perspective, it is essential to develop low complexity sub-optimal scheduling algorithms as  proposed in, e.g.,  \cite{8377145,1705935,5360758,1492685,04668538,m06094252,m07127500,m06884146,m06831724,06095627,m07037316,m06941317,06850064,05456039,05336871,01312486,01664083,04635028,04570215,04299613,05374079,05871799,06504540,06878495,06740115,06725595,05999732,06810624,04784356}. However, due to the lack of analytical- and simulation-based results on the performance of the optimal scheduler, e.g., in terms of sum throughput, it is difficult to evaluate the efficiency of the sub-optimal schemes, unless for small systems. Therefore, it is interesting to derive closed-form expressions for the ultimate performance of optimal schedulers with moderate/large number of transmit antennas and users. Particularly, as shown in the following, these expressions  provide insight into the effect of scheduling on the performance of multi-user MIMO systems, facilitate efficient numerical evaluations for the cases of practical interest, and provide {benchmarks} for evaluation/comparison of different scheduling algorithms.

%%Therefore, as discussed in, e.g., [], a multi-beam satellite network can be mapped into equivalent Multiple-Input Multiple-Output (MIMO) systems
%
%In machine learning, reinforcement algorithms refer to the schemes where dynamic parameter adaptation is performed based on a reward-punishment strategy \cite{reinforcement}. The previous trial(s) being successful, more aggressive parameter settings are risked. On the other hand, the parameters of the upcoming trials are designed more conservatively, if the previous gambling fails. Reinforcement learning differs from the standard \emph{supervised} learning in that the correct input/output pairs are never presented, but a reward-punishment signal is used for parameter adaptation, and the goal is to maximize some notion of the cumulative reward. Due to its generality, the reinforcement algorithm is applied in different fields, such as game theory, control theory, simulation-based optimization and statistics, e.g., \cite{1167350,4445757,4581648,6117774}. However, except some works in these last years, e.g., \cite{4581648,6117774,4278411,4305437,4156378,6542770,6503987}, the reinforcement concept has not been well studied in wireless communication.
%%there are few works utilizing the reinforcement concepts in wireless communication \cite{4581648,6117774,4278411,4305437,4156378}.

In this paper, we study the performance of the scheduling-based multi-user multiple-input-single-output (MISO)  {systems} with large but finite numbers of antennas at the transmitter and single-antenna receivers. The problem is cast in form of optimizing the system sum throughput. The results are obtained for the cases with continuous and bursty communications where in each slot the users may ask for data transmission with different probabilities. Moreover, considering various users' data request probabilities, we maximize the throughput in the cases with different levels of interference cancellation.

%In this paper, we elaborate on the performance of the return-link multi-beam satellite systems utilizing schedulers. The problem is cast in form of optimizing the system sum throughput. The results are obtained for the cases with bursty users' data requests and we take different kinds of interference, namely, the ACI and the CCI, into account. Moreover, considering various users' data request probabilities, we maximize the throughput in different scenarios with and without successive interference cancellation.

The contributions of the paper are threefold. 1) We derive quasi-closed-form expressions for the maximum achievable throughput of the scheduler-based  networks in the cases with different interference cancellation levels (Theorems 1-4, Corollary 1). The derived expressions can be utilized as the benchmarks for the evaluation of different scheduling schemes. Moreover, 2) we develop an efficient scheduling scheme based on GAs. With the proposed GA-based scheduler, the data requesting users are dynamically scheduled in each slot such that the network throughput is optimized. In the meantime, various scheduling rules, e.g., proportional fairness, can be effectively optimized by the proposed algorithm. Finally, 3) we evaluate the effect of different parameters such as the power amplifiers' efficiency, number of antennas/users, different channel and precoding models, scheduling delay and users' data request probabilities on the network performance. Particularly, we use the recent results of \cite{5452208} on the achievable rates of finite block-length codes to analyze the throughput in the cases with short packets, and evaluate the effect of the codeword length on the system performance.

Our paper is different from the state-of-the-art literature because the proposed GA-based scheduler and the derived quasi-closed-form expressions of the throughput have not been presented before.
%As opposed to \cite{7801976} considering antennas selection, we study scheduling-based MIMO networks with different precoding/channel models and analyze the system performance analytically.
Also, we perform finite block-length analysis of the network and present discussions on the effect of power amplifiers, different channel models as well as the cost of iterative scheduling schemes, which have not been considered by, e.g., \cite{1705935,5360758,1492685,m06094252,m07127500,m06884146,m06831724,06095627,m07037316,m06941317,06850064,05456039,05336871,01312486,01664083,04635028,04668538,04570215,04299613,05374079,05871799,06504540,06878495,06740115,06725595,05999732,06810624,04784356}. Also, compared to our preliminary results in \cite{8377145}, this paper performs a deep analysis of the effect of various parameters, percoding schemes and GA-based scheduler on the system performance, and derives closed-form expressions for the network maximum sum rate in different scenarios. Finally, note that, while the paper concentrates on the multi-user MISO networks, the analytical results and the developed scheduler are useful for the analysis of, e.g., return-link multi-beam satellite systems \cite{6127644} as well.

The numerical and the analytical results indicate that the proposed GA-based scheduler reaches (almost) the same performance as in the optimal exhaustive search-based scheduler, with substantially less implementation complexity. %The interference cancellation leads to substantial throughput improvements, particularly when the users' data request probability or the signal-to-noise ratio (SNR) increases.
With different precoders and channel models, there are mappings between the performance of scheduler-based large-but-finite multi-user MISO networks, in the sense that the sum throughput achieved with different precoding schemes/channel models are the same as long as specific parameters are set appropriately. Taking the scheduling
delay into account, the maximum end-to-end throughput is reached by dedicating a small fraction of the packet period
to finding a sub-optimal scheduling rule and using the rest of the packet for data transmission. Moreover, the system throughput is sensitive to the length of short packets while its sensitivity to the packets length decreases for long packets. Finally, the power amplifiers' inefficiency affects the network performance remarkably. For example, consider a $40\times60$  network with continuous communications, no interference cancellation and the typical parameter settings of power amplifiers. Then, with a total consumed power $56$ dBm, the throughput reduces by $50\%$ when the power amplifiers' efficiency decreases from $75\%$ to $25\%.$

The paper works on both the algorithmic and the theoretical analysis of scheduler-based networks, and interestingly these results match with high accuracy. For this reason, depending on the reader's point of interest, there are different efficient ways of reading the paper. For a reader interested in scheduling algorithms, an efficient way to read this paper is to first read Sections II-IV, where the system model, the problem formulation and the proposed GA-based algorithm are described, respectively, and then follow Section VI analyzing the simulation results/algorithm performance. On the other hand, the mathematical analysis of the optimal schedulers is presented in Section V. Thus, a reader with theoretical background may skip Section IV, and read Sections II, III, V and VI in which the system model, the problem formulation, the quasi-closed-form expressions for the maximum throughput and the simulation results are presented, respectively.
\vspace{-0mm}
\section{System model}
We concentrate on the downlink of a multiuser MISO  network with $M$ transmit antennas. Also, we consider bursty communications setup with a maximum of $\tilde N\ge M$ single-antenna users, where in each slot each user may ask for new data with probability $\alpha.$ Here, $\tilde N$ can be the total number of users registered in a base station and is used to derive the expected sum rate in bursty communications model. With this setup, in each time slot, with probability $\Pr(N)={\tilde N \choose N} \alpha^N (1-\alpha)^{\tilde N-N},$ $N$  out of $\tilde N$ users may ask for data. Then, the transmitter schedules $\breve N\le M$ users out of the $N$ data requesting ones and sends them their corresponding messages. Note that setting $\alpha=1$ represents the cases with continuous communications where all $\tilde N$ users are always asking for new information.
%To reduce the implementation complexity, we consider no precoding scheme. In the meantime, it is straightforward to consider different precoding schemes in the proposed GA-based scheduler.
In this way, serving $\breve N\le M$ users in a time slot, the received signal is represented as
\vspace{-0mm}
\begin{align}\label{eq:channelmodel}
\vspace{-0mm}
{\bf{y}}(t) = {\bf{H}}(t){\bf{V}}(t){\bf{s}}(t) + {\bf{z}}(t).
\vspace{-0mm}
\end{align}
Here, ${\bf{H}}(t)\in \mathcal{C}^{\breve N\times M}$ is the fading matrix with the $(i,j)$-th element given by $H_{i,j}(t)=d_{i,j}^{\zeta_{i,j}}h_{i,j}(t)$ where $d_{i,j}$ is the distance between the receiver $i$ and antenna $j$, $\zeta_{i,j}$ is given by the path loss exponent and ${h}_{i,j}(t)\in \mathcal{C}$ denotes the small scale fading. Then, ${\bf{s}}(t)\in \mathcal{C}^{\breve N\times 1}$ denotes the transmitted message vector, ${\bf{V}}(t)\in \mathcal{C}^{M\times \breve N}$ is the precoding matrix  and ${\bf{z}}(t)\in \mathcal{C}^{\breve N\times 1}$ denotes the independent and identically distributed (IID) zero-mean complex Gaussian noise vector with normalized variance. Also, depending on the precoder type, different power normalization constraints can be considered for the precoder. {The proposed GA-based scheduler is applicable for different channel/path loss models. In Section V, however, we derive quasi-closed-form expressions for the achievable throughput in the cases with no path loss. Finally, }
to simplify the presentation, we drop the time index $t$ in the following discussion.

%where $\mathbf{z}\in\mathcal{CN}^{N\times 1}$ is the additive white Gaussian noise, $\mathbf{x}\in\mathcal{C}^{M\times 1}$ denotes the transmitted signal and $\mathbf{y}\in\mathcal{C}^{N\times 1}$ is the corresponding received signal. Moreover, $\mathbf{H}\in\mathcal{C}^{N\times M}$ represents the channel matrix.

%The proposed GA-based scheduler is applicable for different channel models (see Fig. 9). Then, in Theorems 1-4 the ultimate performance of the optimal schedulers are presented for the cases with Rayleigh fading channel conditions, i.e., $\mathbf{H}\in\mathcal{CN}^{N\times M}$. However, as demonstrated in Corollary 1, there are mappings between the sum throughput of MIMO  networks with different fading conditions/precoding schemes, if specific parameters are adapted appropriately.

% (see Corollary 1 for discussions and Fig. 9 for performance analysis in different channel models).

We study quasi-static conditions where the channel coefficients remain constant during the channel coherence time and then change to other values based on their probability density functions (PDFs). Also, we denote the PDF and the cumulative distribution function (CDF) of the random variable $\Delta$ by $f_\Delta$ and $F_\Delta$, respectively. The proposed scheduling algorithm is applicable for the cases with different levels of partial CSI at the transmitter and receivers. In Section V  deriving closed-form expressions for the network sum rate and the simulation figures, however, the channel coefficients are assumed to be known by the transmitter and the receivers. This is an acceptable assumption in quasi-static conditions with larger coherence time of the channel compared to the data transmission periods. As a motivation for this model, consider the long-term evolution (LTE) standard; as discussed in \cite{6477555}, each transmission slot is 0.5 ms in the LTE standard. Also, for systems operating at a carrier frequency around 2.5 GHz and in the case that a receiver is moving with a speed of 2 km/h (walking speed) the coherence time is equal to 200 ms \cite{6477555}. This coherence time is 400 times larger than the time slot duration in LTE\footnote{The coherence time may be shorter in, e.g., millimeter wave (MMW) communications, with high carrier frequencies. However, as future large-but-finite MIMO systems are envisioned to operate in time division duplex (TDD) mode, the channel reciprocity will help to reduce the overhead of CSIT acquisition. Also, with MMW the slot duration can also become shorter such that the coherence time is still much larger than the slot duration.}. Thus, the quasi-static condition can properly model the channel characteristics in the cases with stationary and slow-moving users on which we concentrate.
%(also, see Section VI for discussions on temporally-correlated channel conditions).
Finally, the transmitter is supposed to serve as many users as possible in each time slot.
%In Sections III, we develop the GA-based scheduling technique. Then, we use the algorithm to evaluate the system performance for different users' data request probabilities and channel models. Later, in Section IV, we derive the maximum achievable throughput of the network considering Rayleigh-fading conditions.
\vspace{-0mm}
\section{Problem Formulation}
Let us denote the set of users asking for data in a time slot by $\mathcal{X}\subseteq \{1,\ldots,\tilde N\}$, and the cardinality of $\mathcal{X}$ by $C_\mathcal{X}$. Then, the sum throughput, averaged over many time slots, is given by
\vspace{-0mm}
\begin{align}\label{eq:eta1}
\vspace{-0mm}
\eta&= \sum_{\forall \mathcal{X}}{\left(\Pr( \mathcal{X})\mathbb{E}\{R(\mathbf{H}|\mathcal{X})\}\right)}\mathop  = \limits^{(a)}
\sum_{N=1}^{\tilde N}{\left(\Pr(N)\mathbb{E}\{R(N)\}\right)}.
\vspace{-0mm}
\end{align}
Here,
\vspace{-0mm}
\begin{align}
\Pr(\mathcal{X})=\alpha^{C_\mathcal{X}}(1-\alpha)^{\tilde N-C_\mathcal{X}}
 \end{align}
is the probability that specific users $n\in\mathcal{X}$ ask for data transmission (and the rest remain silent). Also, the probability that $N$ users ask for data transmission, independently of the users' indices, is given by
\vspace{-0mm}
\begin{align}
\Pr(N)={\tilde N \choose N} \alpha^N (1-\alpha)^{\tilde N-N},
\end{align}
%with ${n \choose k}$ being the ``\emph{n choose k}'' operator.
In (\ref{eq:eta1}), $(a)$  holds in the cases with identical long-term channel statistics of the users, on which we concentrate in the simulations. Then, denoting the expectation operator by $\mathbb{E}\{\cdot\}$, $\mathbb{E}\{R(\mathbf{H}|\mathcal{X})\}$ stands for the expected achievable throughput given the  data requesting users $n\in\mathcal{X}$, with expectation over all possible channel realizations $\mathbf{H}$.
%Considering a maximum of $\tilde N$ users, the network sum throughput is given by
%\begin{align}\label{eq:eta1}
%\vspace{-0mm}
%\eta= \sum_{\forall \mathbf{\hat H}^n}{\Pr( \mathbf{\hat H}^n)E\{R(\mathbf{\hat H}^n)\}}.
%\vspace{-0mm}
%\end{align}
%Here, $\Pr(\mathbf{\hat H}^n)$ is the probability of scheduling $n\le \tilde N$ specific users associated with the channel $\mathbf{\hat H}^n$. Also, denoting the expectation operator by $E\{.\}$, $E\{R(\mathbf{\hat H}^n)\}$ stands for the expected achievable throughput in the cases with the channel matrix $\mathbf{\hat H}^n$. Moreover, the summation is on all possible channel realizations. Note that, with the data request probability $\alpha$ for each user, we have
%\begin{align}\label{eq:PrHn}
%\vspace{-0mm}
%\Pr(\mathbf{\hat H}^n)= \alpha^n (1-\alpha)^{\tilde N-n}.
%\vspace{-0mm}
%\end{align}
%Also, the probability that, independently of the users' indices, $n$ users request for data transmission (and the rest remain silent) is given by
%\begin{align}\label{eq:Prn}
%\vspace{-0mm}
%\Pr(n)= {\tilde N \choose n} \alpha^n (1-\alpha)^{\tilde N-n}
%\vspace{-0mm}
%\end{align}
%%if $n$ users are scheduled (and the rest remain silent).
%where ${n \choose k}$ is the ``\emph{n choose k}'' operator.
With $M$ transmit antennas, all users asking for data are scheduled if $N\le M,$ i.e., the number of data requesting users is less than the number of transmit antennas. On the other hand, if $N>M$ users ask for data transmission, the scheduler selects the {best} $M$ users out of $N$, such that the throughput is maximized. {That is, $\breve N=\min(N,M)$ in (\ref{eq:channelmodel}).}

Assuming the cases with a power-normalized precoder $\bf{V}$, $M$ transmitting antennas and serving, e.g., $M$ users, the achievable rate terms ${R(\mathbf{H}|\mathcal{X})}$ and ${R(N)}$ are respectively obtained by
\vspace{-0mm}
\begin{align}\label{eq:expectedwithout}
\vspace{-0mm}
R(\mathbf{H}|\mathcal{X})= \sum_{\forall i\in\mathcal{X}}{\log\bigg(1+\frac{\frac{P}{\breve N} g_{i,i}}{\frac{P}{\breve N}\sum_{\forall j\in\mathcal{X},j\ne i}{ g_{i,j}}+1}\bigg)},
\vspace{-0mm}
\end{align}
and
\vspace{-0mm}
\begin{align}\label{eq:expectedwithout2}
\vspace{-0mm}
R(N)= \sum_{i=1}^N{\log\bigg(1+\frac{\frac{P}{\breve N} g_{i,i}}{\frac{P}{\breve N}\sum_{j=1,j\ne i}^{N}{ g_{i,j}}+1}\bigg)},
\vspace{-0mm}
\end{align}
nats per channel use (npcu), where $g_{i,j}$ is the $(i,j)$-th element of the matrix ${\mathbf{G}}=|{\mathbf{H}}\bf{V}|^2$. Also, $P$  is the total transmission power. Thus, because the noise variance is set to 1, $P$ (in dB, $10\log_{10}P$) represents the signal-to-noise ratio (SNR) at the transmitter as well.

{As the other extreme case, one can consider the cooperative upper bound of MIMO channels, which is achieved under the assumption that the users can cooperate with each other in decoding their received signals, leading to \cite[Eq. (7)]{6127644}, \cite{1421925}}
\begin{align}\label{eq:expectedwith}
\vspace{-0mm}
R(N)^\text{CO}= \log\det\left(\mathbf{I}_{N}+\frac{P}{\breve N}\mathbf{ H}\mathbf{ H}^\text{h}\right).
\vspace{-0mm}
\end{align}
Here, $\mathbf{ H}^\text{h}$ denotes the Hermitian transpose of $\mathbf{H}$ and $\mathbf{I}_N$ is the $N\times N$ identity channel matrix,  and the result is achieved under the assumption of perfect CSI at the receiver. {Also, (\ref{eq:expectedwith}) is based on the fact that with users' cooperation the system becomes equivalent to a point-to-point MIMO connection, and the sum rate is achieved through the use of successive interference cancellation.} Note that, depending on the number of antennas and users, {the cooperation} may not be possible and the rate (\ref{eq:expectedwith}) is not practically achievable.
% because it needs cooperation between all users.
%For this reason, we mainly concentrate on the cases with no interference cancellation.
However, it is interesting to derive the system performance in the cases with {users' cooperation} because 1) it is a benchmark for the ultimate potential gains of the interference cancellation in MIMO networks and 2) as seen in the following, the performance of the zero-forcing based scheme is close to the one obtained via (\ref{eq:expectedwith}). For this reason, Theorem 4, presented in the sequel, determines the maximum achievable throughput of the network in the cases with {users' cooperation}. Moreover, the mathematical techniques of Theorem 4 can be supportive for different analysis of MIMO system with large-but-finite number of antennas. Finally, Theorems 1-3 and Corollary 1 analyze the system performance for the cases with no interference cancellation and different precoding schemes, respectively.

In this perspective, the scheduling problem is simplified to finding the optimal, in terms of throughput, configuration among the sub-matrices  of the matrix $\mathbf{ H}$ in each slot. Indeed, the optimal set of users can be selected via exhaustive search in the cases with few antennas/users. However, with large networks, which {are} of interest in the next generation of wireless networks, we need to design efficient algorithms to derive sub-optimal scheduling rules with low complexity.

\vspace{-0mm}
\section{Genetic Algorithm-based Scheduling}

{In this paper, we propose a scheduling scheme based on GAs \cite{7801976,1705935,1492685,04668538,5360758}. The proposed scheme is} explained in Algorithm 1. In words, the algorithm is based on the following procedure. If $N\le M$ users ask for data, serve all of them. Otherwise, do the following. With a given precoding scheme, start the algorithm by selecting $K$ possible channel assignments. Each channel assignment corresponds to a selected set of users. In each iteration, we determine the best matrix, referred to as the \emph{queen}, that leads to the highest throughput, compared to the other considered matrices. Then, we keep the queen for the next iteration and create $J<K$ matrices around the queen. This is achieved by applying small modifications into the queen (For instance, by changing a few number of users in the set of users associated with the queen). Finally, in each iteration $K-J-1$ sets of user assignments are selected randomly and the iterations continue for $N_\text{it}$ times considered by the designer. Running all considered iterations, the queen is returned as the scheduling rule of the current network realization.
%Moreover, the rest of $M-J$ remaining matrices of the next generation are created randomly.
%Note that in the cases with $n>br$ the channel selection is equivalent to select a sub-channel matrix $\mathbf{H}^{br}$ from a parent matrix of size $\mathbf{H}^n.$ On other hand, considering $n\le br$, the channel selection is equivalent to proper selection of a matrix $\mathbf{H}^n$ and inserting $(br-n)$ zero columns into the matrix (related to non-selected bearers) to reach the channel matrix $\mathbf{H}^{br}$.
The throughput is achieved by averaging on the achievable rates over many channel realizations.
\begin{algorithm}
\caption{GA-based Scheduling Algorithm}
%\begin{algorithmic}
In each slot with $N$ users asking for data, do the followings:
\begin{itemize}
\item \emph{If $N>M$ }
\begin{itemize}
\item[I.] Consider $K$ sets of $M$ users and for each set create the channel matrix. Consequently, $K$ associated matrices ${\mathbf{ H}^{k}}, k=1\ldots,K,$ are created.
   % {\pi ^l} = \sum\limits_{i = 1}^{{2^N}} {p_i^l\sum\limits_{k = i + 1}^{{2^N}} {{p_{k|i}}\,} }  + \sum\limits_{i = 1}^{{2^N}} {{p_{1|i}}p_i^l}  \le {P_{{\rm{outage}}}},\,\forall l = 1...L$ in which we have $p_i^l = \int\limits_{{g_{i - 1,l}}}^{{g_{i,l}}} {{f_G}(g)dg}$.
\item[II.] For each matrix ${\mathbf{ H}^{k}}, k=1\ldots,K$, use (\ref{eq:expectedwithout}) and the considered precoding scheme to determine the throughput $R({\mathbf{ H}^{k}})$.
%\item[II.] For each vector, determine the last retransmission power $T_{M+1}^j$ according to (15). If $T_{M+1}^j <0$, eliminate the $j$-th vector.
%\item[III.] Find the vectors throughput
%\begin{itemize}
%  \item Find the probability coeficients $\beta_{n,k}$'s and $p_n$'s based on the joint pdf.
%  \item Find the optimal transmission powers according to (8) and (10).
%  \item Determine the throughput ${\hat \eta}^j$ based on the transmission powers and (7).
%\end{itemize}
\item[III.] Find the matrix which results in the highest throughput, referred to as the queen, i.e., ${\mathbf{ H}^{i}}$ where $R({\mathbf{ H}^{k}}) \le R({\mathbf{ H}^{i}}),\,\forall k = 1,\ldots,K$.
\item[IV.] ${\mathbf{ H}^{1}} \leftarrow {\mathbf{ H}^{i}}$.
\item[V.] Generate {$J < K$} matrices ${\mathbf{ H}^{j,\text{new}}},\,j = 1,\ldots,J,$ around ${\mathbf{ H}^{1}}$. These matrices are generated by small changes in the queen; for instance, by changing a number of the users in the queen.
\item[VI.] ${\mathbf{ H}^{j+1}} \leftarrow {\mathbf{ H}^{j,\text{new}}},\,j = 1,\ldots,J$.
\item[VII.] Regenerate the remaining matrices ${\mathbf{ H}^{j}},j = J + 2,\ldots,K,$ randomly with the same procedure as in Step I.
\item[VIII.] Go to II. and continue the procedure for $N_\text{it}$ iterations where $N_\text{it}$ is the number of iterations considered by the designer.

\end{itemize}
\item{\emph{else if $N\le M$ }}

 Schedule all data requesting users and calculate the throughput based on (\ref{eq:expectedwithout}).
\end{itemize}
Return the queen as the scheduling rule of the current slot.
\end{algorithm}

Considering the proposed GA-based algorithm, it is interesting to note that:
\begin{itemize}
  \item The algorithm is generic in the sense that it can be implemented for different channel models, precoding schemes and objective functions.
  \item As opposed to exhaustive search-based methods, the proposed algorithm implies $KN_\text{it}$ trials which, depending of the considered parameter settings, can be considerably low. Particularly, as demonstrated in Section VI, the proposed approach reaches (almost) the same throughput as in the optimal (exhaustive-search) scheduler with few iterations.
       %(also, see Fig. 10b on the cost of scheduling delay).
      %Therefore, the algorithm is reasonably fast and it can be implemented for large networks (also, see Fig. 10b on the cost of scheduling delay).
  \item Because of step VII. of the algorithm, where $K-J-1$ random channel assignments are checked in each iteration, Algorithm 1 mimics the exhaustive search and reaches the globally optimal selection rule if $N_\text{it}\to\infty$.
  \item {As opposed to typical GAs, we do not use the crossover operation because the proposed algorithm works very well with no need for the additional complexity of the crossover operation. However, it is straightforward to include the crossover into the proposed algorithm where, for instance, the queen and the next best solutions are combined to generate new possible solutions.}
  \item {For simplicity, we presented Algorithm 1 for the cases with perfect CSI available at the transmitter. However, the proposed algorithm is well applicable for the cases with different levels of partial CSI at the transmitter. With imperfect CSI, we follow the same approach as in Algorithm 1, except that the precoding matrices are designed based on the partial CSI available at the transmitter which will affect the sum throughput (\ref{eq:expectedwithout}) correspondingly.}
      %That is, the proposed scheme is optimal when the number of iterations increases asymptotically.
\end{itemize}
Performance analysis of the proposed algorithm is studied in Figs. 6-12 and Table I.

%\begin{align}\label{eq:eta}
%\vspace{-0mm}
%\eta= \sum_{n=1}^{\tilde N}{\Pr(n)E\{R(n)\}}.
%\vspace{-0mm}
%\end{align}
%Here, $\Pr(n)$ is the probability that $n$ users request for data transmission (and the rest remain silent). Also, $E\{R(n)\}$ denotes the expected achievable throughput in the cases with $n$ users with $E\{.\}$ standing for the expectation operator.
\vspace{-0mm}
\section{Throughput Analysis in Rayleigh Fading Channels}
Here, we consider Rayleigh fading model and derive approximations/bounds for the ultimate performance of optimal schedulers. Particularly, Theorems 1-3 and 4 present quasi-closed-form expressions for the maximum throughput of the scheduler-based scheme in the cases {with no interference cancellation and users' cooperation}, respectively. Then, Corollary 1 and Subsection V.C extend the results to the cases with different precoding/fading models and short packets, respectively.
\subsection{Throughput with no Interference Cancellation}
{With multiple antennas at the transmitter, different precoding schemes are normally applied by the transmitter, which increase the users' received signal-to-interference-plus-noise ratio (SINR). In this section, however, we start the discussions by assuming no precoding at the transmitter. Although it leads to low achievable rates for the users at high SNRs, the performance analysis with no interference cancellation is important because 1) it provides a lower bound for the performance of precoding-based schemes, and 2) the same throughput term as in (\ref{eq:expectedwithout}) is applicable for different multiuser networks, such as point-to-point spectrum sharing networks scheduled by a central unit. Also, 3) as we show in Fig. 4, the no interference cancellation scheme is of interest at low SNRs, because it provides relatively good performance compared to the cases using zero-forcing precoder/users' cooperation with considerably less implementation complexity.}

Considering no interference cancellation at the receivers, the achievable rate in each time slot is given by (\ref{eq:expectedwithout2}) with ${\mathbf{G}}=|{\mathbf{H}}|^2$. Here, we consider two cases with $N\ge M$ and $N< M$ and find the expected achievable rate as given in (\ref{eq:Rngebr2}) and (\ref{eq:Rnlebr}), respectively. Then, the results of (\ref{eq:Rngebr2}) and (\ref{eq:Rnlebr}) are combined to determine the throughput as presented in Theorem 1.
%The performance analysis with no interference cancellation is important because 1) it provides a lower bound for the performance of precoding-based schemes, and 2) the same throughput term as in (\ref{eq:eqtheorem1}) is applicable for different multiuser networks. Also, 3) as we show in Fig. 4, the no interference cancellation scheme is of interest at low SNRs, because it provides relatively good performance compared to the cases with successive interference cancellation.

%\textcolor{red}{ta inja}

In each time slot, if $N> M$ users ask for data transmission, there are ${N \choose {M}}$ combinations of possible user selections and the optimal scheduler picks the best combination such that the throughput is maximized. Therefore, with $N> M,$ we have
\vspace{-0mm}
\begin{align}\label{eq:Rngebr1}
\vspace{-0mm}
\mathbb{E}\{R(N)^\text{no-IC}|N> M\}&= \mathbb{E}\{Z^{N,M}\}=\int_0^\infty{zf_{Z^{N,M}}(z)\text{d}z}\nonumber\\&\mathop  = \limits^{(b)}\int_0^\infty{\left(1-F_{Z^{N,M}}(z)\right)\text{d}z}.
\vspace{-0mm}
\end{align}
Here, $(b)$ is obtained by partial integration and $Z^{N,M}$ is the random variable defined as
\vspace{-0mm}
\begin{align}\label{eq:Zdef}
Z^{N,M}\doteq\mathop {\max }\limits_{n=1,\ldots, {N \choose {M}}}\left\{Z_n^{N,M}\right\},
\end{align}
%\textcolor{red}{ta inja}
with
\vspace{-0mm}
\begin{align}\label{eq:Zidef}
&Z_n^{N,M}=\nonumber\\&\left\{\sum_{i=1}^{M}{\log\left(1+u_i^{N,M}\right)}\bigg| \text{selecting the } \,n\text{-th combination}\right\},\nonumber\\&u_i^{N,M}\doteq\frac{\frac{P}{M} g_{i,i}^N}{1+\frac{P}{M}\sum_{j=1,\ldots,M,j\ne i}{ g_{i,j}^N}}.
\end{align}
%Note that, for IID Gaussian channels, the PDFs of random variables $ g_{i,j}^N$ and $\chi_i^{N,M}=\sum_{j=1,\ldots,M,j\ne i}{ g_{i,j}^N}$ are given by $f_{ g_{i,j}^N}(x)=e^{-x}$ and $f_{\chi_i^{N,M}}(x)=\frac{1}{\Gamma(M-1)}x^{M-2}e^{-x},x\ge 0,$ respectively, where $\Gamma(x)=\int_0^\infty{t^{x-1}e^{-t}\text{d}t}$ denotes the Gamma function.
{Note that the PDF of the random variable $g_{i,j}^N=|h_{i,j}^N|^2$ is given by $f_{g_{i,j}^N}(x)=e^{-x}$ because $h_{i,j}^N$ is Rayleigh distributed. Also, considering the random variable $S(n)=\sum_{i=1}^{n}g_i$, with $f_{g_i}(x)=e^{-x}$, we have $F_{S(1)}(x)=1-e^{-x}$ and}
\begin{align}\label{eq:Eqcdfchi}
F_{S(n)}(x)&=\Pr\left(S(n-1)+g_n\le x\right)=\int_0^x{e^{-t}F_{S(n-1)}(x-t)\text{d}t}\nonumber\\&=1-e^{-x}\sum_{i=1}^{n-1}\frac{x^i}{i!}, n\ge 2,
\end{align}
{which leads to the PDF $f_{S(n)}(x)=\frac{x^{n-1}}{\Gamma(n)}e^{-x}.$ Thus, using (\ref{eq:Eqcdfchi}), the PDF of the random variable $\chi_i^{N,M}=\sum_{j=1,\ldots,M,j\ne i}{ g_{i,j}^N}$ is given by $f_{\chi_i^{N,M}}(x)=\frac{1}{\Gamma(M-1)}x^{M-2}e^{-x},x\ge 0.$} In this way, from (\ref{eq:Zidef}), we have
\vspace{-0mm}
\begin{align}\label{eq:CDFUim}
&F_{u_i^{N,M}}(u)=\Pr\left(u_i^{N,M}\le u\right)\nonumber\\&=\int_0^\infty{\frac{1}{\Gamma(M-1)}x^{M-2}e^{-x}\Pr\left( g_{i,i}^N\le \frac{Mu}{P}\left(\frac{P}{M}x+1\right)\right)\text{d}x}
\nonumber\\&\mathop  = \limits^{(c)}\int_0^\infty{\frac{1}{\Gamma(M-1)}x^{M-2}e^{-x}\left(1-e^{-\frac{Mu}{P}\left(\frac{P}{M}x+1\right)}\right)\text{d}x}
\nonumber\\&=1-\frac{1}{\Gamma(M-1)}e^{-\frac{Mu}{P}}\int_0^\infty{x^{M-2}e^{-(1+u)x}\text{d}x}\nonumber\\&
\mathop  = \limits^{(d)}1-\frac{e^{-\frac{Mu}{P}}}{(1+u)^{M-1}},
\end{align}
where $(c)$ is obtained by $F_{ g_{i,i}^N}(x)=1-e^{-x},x\ge0,$ for IID Gaussian channels. Also, $(d)$ comes from some manipulations and the definition of the upper {incomplete} Gamma function $\Gamma(s,x)=\int_x^\infty{t^{s-1}e^{-t}\text{d}t},$ $\Gamma(s)=\Gamma(s,0).$

Using (\ref{eq:Zdef}), the CDF of the random variable $Z^{N,M}$ is found as
\begin{align}\label{eq:CDFFz1}
F_{Z^{N,M}}(z)=\bigg(F_{Z_n^{N,M}}(z)\bigg)^{N \choose {M}}.
\end{align}
Therefore, the final step to find $F_Z(z)$ and (\ref{eq:Rngebr1}) is to derive $F_{Z_n^{N,M}}(z).$ Using (\ref{eq:Zidef}) and the central limit Theorem (CLT) and for moderate/large values of $M$, which is our range of interest, the variable $Z_n^{N,M}$ converges to the Gaussian variable $\mathcal{Z}\sim\mathcal{N}({M}\mu,{M}{\sigma^2})$ with $\mu$ and $\sigma^2$ given by
\begin{align}\label{eq:eqmu}
\mu&=E\left\{\log\left(1+u_i^{N,M}\right)\right\}=\int_0^\infty{\log(1+x)f_{u_i^{N,M}}(x)\text{d}x}\nonumber\\&\mathop  = \limits^{(e)}\int_0^\infty{\frac{1-F_{u_i^{N,M}}(x)}{1+x}\text{d}x}=\int_0^\infty{\frac{e^{\frac{-Mx}{P}}}{(1+x)^{M}}\text{d}x}\mathop  = \limits^{(f)}e^{\frac{M}{P}}\text{E}_{M}\left(\frac{M}{P}\right),
\end{align}
and $\sigma^2=\rho-\mu^2$ with
\vspace{-0mm}
\begin{align}\label{eq:eqsigma}
&\rho=E\left\{\log\left(1+u_i^{N,M}\right)^2\right\}=\int_0^\infty{\log(1+x)^2f_{u_i^{N,M}}(x)\text{d}x}
\nonumber\\& \mathop  = \limits^{(g)}2\int_0^\infty{\frac{\log(1+x)}{1+x}\left(1-F_{u_i^{N,M}}(x)\right)\text{d}x}
\nonumber\\&=2\int_0^\infty{\frac{\log(1+x)e^{-\frac{Mx}{P}}}{(1+x)^{M}}\text{d}x}
\nonumber\\&\mathop  \simeq \limits^{(h)}2\int_0^\infty{\log(1+x)e^{-\left(\frac{M}{P}+M\right)x}\text{d}x}\mathop  = \limits^{(i)}\frac{2Pe^{\frac{M}{P}+M}}{M+PM}\text{E}_1\left(\frac{M}{P}+M\right),
\end{align}
respectively. Here, $\text{E}_n(x)=\int_1^\infty{\frac{e^{-xt}}{t^n}\text{d}t}$ is the $n$-th order exponential integral function. Also, $(e)$ is obtained by partial integration and $(f)$ follows from some manipulations and the definition of the exponential function. In (\ref{eq:eqsigma}), $(g)$ comes from partial integration. Then, $(h)$ is based on the approximation $\frac{1}{(1+x)^n}\simeq e^{-nx},\forall n>0,$ which is tight for moderate/large values of $M$ on which we concentrate. Finally, $(i)$ follows from straightforward manipulations and using the definition of the exponential integral function.

In this way, considering the error function $\text{erf}(x)=\frac{2}{\sqrt{\pi}}\int_0^x{e^{-t^2}\text{d}t}$ and the CDF of the Gaussian variables, we have
\begin{align}\label{eq:mathcalzcdf}
F_{\mathcal{Z}}(z)=\frac{1}{2}\left(1+\text{erf}\left(\frac{z-M\mu}{\sqrt{2M\sigma^2}}\right)\right),
\end{align}
and, from (\ref{eq:CDFFz1}),
\begin{align}\label{eq:CDFFz2}
F_{Z^{N,M}}(z)=\left(\frac{1}{2}\right)^{N \choose {M}}\left(1+\text{erf}\left(\frac{z-M\mu}{\sqrt{2M\sigma^2}}\right)\right)^{N \choose {M}}.
\end{align}
Thus, the expected term (\ref{eq:Rngebr1}) is given by
\begin{align}\label{eq:Rngebr2}
\vspace{-0mm}
&E\left\{R(N)^\text{no-IC}|N> M\right\}=\mathcal{Q}^\text{no-IC}(N,M,P),\nonumber\\&\mathcal{Q}^\text{no-IC}(N,M,P)\nonumber\\&=\int_0^\infty{\bigg(1-\left(\frac{1}{2}\right)^{N \choose {M}}\left(1+\text{erf}\left(\frac{z-M\mu}{\sqrt{2M\sigma^2}}\right)\right)^{N \choose {M}}\bigg)\text{d}z},
\vspace{-0mm}
\end{align}
which, using $\mu$ and $\sigma$ in (\ref{eq:eqmu}) and (\ref{eq:eqsigma}), can be easily calculated for every given values of $N,M,P$.

Finally, because all users are selected by the scheduler in the cases with $N\le M,$ the expected achievable throughput in that case is found as
\begin{align}\label{eq:Rnlebr}
\vspace{-0mm}
&\mathbb{E}\{R(N)^\text{no-IC}|N\le M\}=E\left\{\sum_{i=1}^N\log\left(1+u_i^{N,M}\right)\right\}
\nonumber\\&=NE\left\{\log\left(1+u_i^{N,M}\right)\right\}\mathop  = \limits^{(j)}Ne^{\frac{N}{P}}\text{E}_{N}\left(\frac{N}{P}\right),
\end{align}
where $(j)$ is obtained with the same procedure as in (\ref{eq:eqmu}). Combining (\ref{eq:Rngebr2}) and (\ref{eq:Rnlebr}), the network throughput in the cases with no interference cancellation is given as summarized in Theorem 1.

\emph{\textbf{Theorem 1.}} Without interference cancellation, the throughput of the scheduler-based network is (approximately) determined as
\begin{align}\label{eq:eqtheorem1}
&\eta^\text{no-IC}=\sum_{N=1}^{M}{{\tilde N \choose N}\alpha^N(1-\alpha)^{\tilde N-N}Ne^{\frac{N}{P}}\text{E}_{N}\left(\frac{N}{P}\right)}\nonumber\\&+\sum_{N=M+1}^{\tilde N}{{\tilde N \choose N}\alpha^N(1-\alpha)^{\tilde N-N}\mathcal{Q}^\text{no-IC}(N,M,P)},
\end{align}
with $\mathcal{Q}^\text{no-IC}(\cdot,\cdot,\cdot)$ given in (\ref{eq:Rngebr2}).\qed
%\vspace{-2mm}

%Note that in the cases with $N\le M$ we can rewrite the results for the cases where $M-N$ antennas are turned off and $N$ users are served by $N$ antennas using $\frac{P}{N}$ per-antenna power. In that case, (\ref{eq:Rnlebr}) is rewritten as $E\{R(N)^\text{no-IC}|N\le M\}=Ne^{\frac{N}{P}}\text{E}_{N}\left(\frac{N}{P}\right),$ and (\ref{eq:eqtheorem1}) is updated correspondingly. However, in our setup with $\tilde N\gg M$ users and  high users' data requesting probabilities, the probability that only $N\le M$ users request for data is very low. That is, while the first term of (\ref{eq:eqtheorem1}) associated with the case $N\le M$ is necessary for the completeness of the analysis, it is by orders of magnitude smaller than the second term of (\ref{eq:eqtheorem1}) corresponding to the cases with $N>M.$ Thus, the first term of (\ref{eq:eqtheorem1}) can be removed, and the approximations are still tight for a broad range of parameter settings.
{Finally,} to find a quasi-closed-form expression for the one-dimensional integration (\ref{eq:Rngebr2}), we use a linearization technique as stated in Theorem 2.

\emph{\textbf{Theorem 2.}} With no interference cancellation, the maximum achievable throughput of the network is approximated as
\begin{align}\label{eq:eqtheorem1approx0}
\eta^\text{no-IC}&\simeq\sum_{N=1}^{M}{{\tilde N \choose N}\alpha^N(1-\alpha)^{\tilde N-N}Ne^{\frac{N}{P}}\text{E}_{N}\left(\frac{N}{P}\right)}\nonumber\\&+\sum_{N=M+1}^{\tilde N}{{\tilde N \choose N}\alpha^N(1-\alpha)^{\tilde N-N}}\tau_{N,M},
\end{align}
with $\tau_{N,M}$ given in (\ref{eq:eqlinearapprox}).
\begin{proof}
As a second order approximation, we can use Theorem 1 and the linearization technique $\mathcal{T}_{N,M}(x)\simeq Y_{N,M}(x)$ where $\mathcal{T}_{N,M}(x)=1-\left(\frac{1}{2}\right)^{N \choose {M}}\left(1+\text{erf}\left(\frac{x-M\mu}{\sqrt{2M\sigma^2}}\right)\right)^{N \choose {M}}$ and
\vspace{-0mm}
\begin{align}\label{eq:eqlinearapprox}
& Y_{N,M}(x)\nonumber\\&=\left\{\begin{matrix}
1 & x\le \tau_{N,M}+\frac{1}{2\zeta_{N,M}}, \\
\frac{1}{2}+\zeta_{N,M}(x-\tau_{N,M}) & x\in\left[\tau_{N,M}+\frac{1}{2\zeta_{N,M}},\tau_{N,M}-\frac{1}{2\zeta_{N,M}}\right],\\
0 & x\ge \tau_{N,M}-\frac{1}{2\zeta_{N,M}},
\end{matrix}\right.\nonumber\\&
\tau_{N,M}=M\mu+Q^{-1}\left(1-\left(\frac{1}{2}\right)^{\frac{1}{{N \choose M}}}\right)\sqrt{M\sigma^2},\nonumber\\&
\zeta_{N,M}=\frac{-1}{2\sqrt{2\pi M\sigma^2}}{N \choose M}e^{-\frac{1}{2}\left(Q^{-1}\left(1-\left(\frac{1}{2}\right)^{\frac{1}{{N \choose M}}}\right)\right)^2},
\end{align}
to rewrite (\ref{eq:eqtheorem1}) as
\vspace{-0mm}
\begin{align}\label{eq:eqtheorem1approx}
\eta^\text{no-IC}&\mathop  \simeq \limits^{(k)}\sum_{N=1}^{M}{{\tilde N \choose N}\alpha^N(1-\alpha)^{\tilde N-N}Ne^{\frac{N}{P}}\text{E}_{N}\left(\frac{N}{P}\right)}\nonumber\\&+\sum_{N=M+1}^{\tilde N}{{\tilde N \choose N}\alpha^N(1-\alpha)^{\tilde N-N}}\int_0^\infty{Y_{N,M}(x)\text{d}x}\nonumber\\&
=\sum_{N=1}^{M}{{\tilde N \choose N}\alpha^N(1-\alpha)^{\tilde N-N}Ne^{\frac{N}{P}}\text{E}_{N}\left(\frac{N}{P}\right)}\nonumber\\&+\sum_{N=M+1}^{\tilde N}{{\tilde N \choose N}\alpha^N(1-\alpha)^{\tilde N-N}}\tau_{N,M},
\end{align}
with $\mu$ and $\sigma^2$ given in (\ref{eq:eqmu}) and (\ref{eq:eqsigma}), respectively, and $Q^{-1}(\cdot)$ representing the inverse $Q$ function. Here, $(k)$ comes from the first order Taylor expansion of $\mathcal{T}_{N,M}(x)$ at point $x=\tau_{N,M}$.
\end{proof}

Setting $\alpha=1$ in (\ref{eq:eqtheorem1}) for continuous communications setups where all users are always asking for data transmission, the network throughput is given by
\vspace{-0mm}
\begin{align}\label{eq:eqcontinuous1}
&\eta^\text{no-IC, continuous}\nonumber\\&=\int_0^\infty{\bigg(1-\left(\frac{1}{2}\right)^{\tilde N \choose {M}}\left(1+\text{erf}\left(\frac{z-M\mu}{\sqrt{2M\sigma^2}}\right)\right)^{\tilde N \choose {M}}\bigg)\text{d}z}.
\end{align}

%Finally, to complete the discussions on the no interference cancellation scenario, Appendix A presents another approximation approach which is useful for the analysis of the networks of small size. However, as the paper concentrates on moderate/large networks, we do not consider small values of $\tilde N, M$ in our numerical evaluations.

In Theorems 1-2, we used the CLT to analyze the performance of the scheduler-based schemes in the cases with large numbers of antennas and users. Theorem 3 completes the discussions on the no interference cancellation scenario, and bounds the sum throughput in the cases with different numbers of antennas/users.

\emph{\textbf{Theorem 3.}} For every number of antennas/users, the network sum throughput is lower-bounded via (\ref{eq:eqalouini})-(\ref{eq:Rngebr22}).

\begin{proof}
From (\ref{eq:CDFUim}), the random variable $u_i^{N,M}$ is dominated\footnote{The random variable $X$ dominates the random variable $Y$ if $F_X(x)\ge F_Y(x),\forall x$ \cite{dominantcdf}.} by the random variable $v_i^{N,M}$ which follows the CDF
\vspace{-0mm}
\begin{align}\label{eq:eqdominantCDF}
F_{v_i^{N,M}}(x)=1-e^{-\left(\frac{M}{P}+M-1\right)x}.
\end{align}
Here, we have used the bound $\frac{1}{(1+x)^n}\ge e^{-nx},\forall n, x\ge0,$ which, considering  (\ref{eq:CDFUim}) and (\ref{eq:eqdominantCDF}), is  tight at low SNRs, but its tightness decreases as the transmission power increases. In this way, we can use the results of \cite{5426254} and (\ref{eq:eqdominantCDF}), to upper-bound the CDF $F_{Z^{N,M}}(z)$ by
\begin{align}\label{eq:eqalouini}
&F_{Z_n^{N,M}}(z)\le F_{V_n^{N,M}}(z),\forall z,\nonumber\\&
F_{V_n^{N,M}}(z)=1-e^{M\left(\frac{M}{P}+M-1\right)} \times\nonumber\\& \mathcal{H}_{1,M+1}^{M+1,0}\bigg[\frac{ 2^{z}}{\log(2)}\bigg(\frac{M}{P}+M-1\nonumber\\&\bigg)^M\bigg |_{(0,1,0),\underbrace{(1,1,\frac{M}{P}+M-1),\ldots,(1,1,\frac{M}{P}+M-1)}_{M\,\, \text{ times}}}^{\,\,\,\,\,\,\,\,\,\,\,\,\,\,\,\,\,\,\,\,\,\,\,\,\,\,\,\,\,\,\,\,\,\,\,\,\,\,\,\,\,\,\,\,\,\,(1,1,0)} \bigg],
\end{align}
where $ \mathcal{H}_{m,n}^{s,r}\big[. \big]$ denotes the generalized upper incomplete {Fox's} H function \cite[Sec. 1.19]{FoxHbook}. In this way, the expected term (\ref{eq:Rngebr1}) can be lower-bounded by the numerical evaluation of
\vspace{-0mm}
\begin{align}\label{eq:Rngebr22}
\vspace{-0mm}
&E\left\{R(N)^\text{no-IC}|N> M\right\}\ge\int_0^\infty{\bigg(1-\left(F_{V_n^{N,M}}(z)\right)^{{ N \choose M}}\bigg)\text{d}z},
\vspace{-0mm}
\end{align}
which, using (\ref{eq:Rnlebr}) and the same arguments as in Theorem 1, lower-bounds the throughput.
\end{proof}
Mathematically, (\ref{eq:Rngebr22}) is applicable for {every value} of $M, N.$ However, for, say $M>5,$ the implementation of the generalized upper incomplete {Fox's} H function is very time-consuming and the tightness of the bound decreases with increasing $M,N$. As a result, (\ref{eq:eqalouini})-(\ref{eq:Rngebr22}) are useful for performance analysis in the cases with small $M,N$'s, while the CLT-based approach of Theorems 1-2 provides accurate performance evaluation for moderate/large networks. Finally, except for Fig. 4 which analyzes small networks, we do not consider small values of $\tilde N, M$ in our evaluations.
\vspace{-0mm}
\subsection{Throughput with Interference Cancellation}
In this subsection, we first analyze the ultimate performance limits of the scheduler-based MIMO  networks in the cases with {users' cooperation}. This is interesting because, as seen in the following, the performance of the optimal scheduler using zero-forcing precoder is close to the one with {users' cooperation} (see Fig. 4).
%Then, Corollary 1 extends the results to the cases with different precoding and fading models.

As illustrated in Section III, from the mathematical perspective the only difference between the cases {with no interference cancellation and users' cooperation} is in their achievable rates given in (\ref{eq:expectedwithout}) and (\ref{eq:expectedwith}), respectively. Therefore, we use (\ref{eq:expectedwith}) and follow the same procedure as in (\ref{eq:Rngebr1})-(\ref{eq:eqtheorem1}) to find the throughput in the cases {with cooperative users}. The details of the analysis are as follows.

With $N>M$ number of data requesting users and interference cancellation, $M$ users are selected by the scheduler and we have
\vspace{-0mm}
\begin{align}\label{eq:Rngebrout1}
\vspace{-0mm}
\mathbb{E}\{R(N)^\text{CO}|N> M\}=\int_0^\infty{\left(1-F_W(w)\right)\text{d}w},
\vspace{-0mm}
\end{align}
with
\vspace{-0mm}
\begin{align}\label{eq:Zdefout}
W^{N,M}\doteq\mathop {\max }\limits_{n=1,\ldots,\, {N \choose {M}}}\left\{W_n^{N,M}\right\},
\end{align}
\vspace{-0mm}
\begin{align}\label{eq:Zidefout}
&W_n^{N,M}=\nonumber\\&\bigg\{\log\det&\left(\mathbf{I}_{M}+\frac{P}{M}\mathbf{ H}\mathbf{ H}^\text{h}\right)\bigg| \text{selecting the }\, n\text{-th combination}\bigg\},
\end{align}
where $\mathbf{ H}$ is the $M\times M$ channel associated with the $n$-th, $n=1,\ldots,{N \choose {M}},$ combination of the scheduled users. Using CLT for the cases with large number of antennas/users, the random variable $W_n^{N,M}$ converges in distribution into the Gaussian random variable $\mathcal{Y}\sim\mathcal{CN}\left(\hat\mu^{M,M},\left(\hat\sigma^{M,M}\right)^2\right)$. Here, $\hat\mu^{N,M}$ and $\hat\sigma^{N,M}$ are the mean and the standard deviation of the random $C=\log\det\left(\mathbf{I}_{N}+\frac{P}{M}\mathbf{ H}\mathbf{ H}^\text{h}\right)$ which can be effectively found via simulations. Also, there are different approximations in the literature for $(\hat\mu^{N,M},\left(\hat\sigma^{N,M}\right)^2)$, e.g.,  \cite[Section II.A]{1327795}
\begin{align}\label{eq:tarokhapp2}
\left(\hat\mu^{N,M},\left(\hat\sigma^{N,M}\right)^2\right)=(MP,P^2),
 \end{align}
at low/moderate SNRs. In this way, following the same arguments as in (\ref{eq:CDFFz2}), we have
\begin{align}\label{eq:CDFFW2}
F_{W^{N,M}}(z)=\left(\frac{1}{2}\right)^{N \choose {M}}\left(1+\text{erf}\left(\frac{z-\hat\mu^{N,M}}{\sqrt{2}\hat\sigma^{N,M}}\right)\right)^{N \choose {M}}.
\end{align}
Thus, the expected term (\ref{eq:Rngebrout1}) is found as
\begin{align}\label{eq:withRngebr2}
\vspace{-0mm}
&E\left\{R(N)^\text{CO}|N> M\right\}=\mathcal{Q}^\text{CO}(N,M,P),\nonumber\\&\mathcal{Q}^\text{CO}(N,M,P)=\nonumber\\&\int_0^\infty{\bigg(1-\left(\frac{1}{2}\right)^{N \choose {M}}\left(1+\text{erf}\left(\frac{z-\hat\mu^{N,M}}{\sqrt{2}\hat\sigma^{N,M}}\right)\right)^{N \choose {M}}\bigg)\text{d}z}.
\vspace{-0mm}
\end{align}
Also, the expected rate in the cases with $N\le M$ number of data requesting users is given by
\begin{align}\label{eq:RnlebrIC}
\vspace{-0mm}
&E\left\{R(N)^\text{CO}|N\le M\right\}=E\left\{\log\det\left(\mathbf{I}_N+\frac{P}{N}\mathbf{H}\mathbf{ H}^\text{h}\right)\right\}=\hat\mu^{N,N},
\end{align}
which can be found via the approximation scheme of (\ref{eq:tarokhapp2}) or by numerical evaluations. Thus, {with users' cooperation the} throughput is found as summarized in Theorem 4.

\emph{\textbf{Theorem 4.}} {With users' cooperation}, the throughput of the scheduler-based network is (approximately) obtained by
\begin{align}\label{eq:eqtheorem2}
&\eta^\text{CO}=\sum_{N=1}^{M}{{\tilde N \choose N}\alpha^N(1-\alpha)^{\tilde N-N}\hat\mu^{N,N}}\nonumber\\&+\sum_{N=M+1}^{\tilde N}{{\tilde N \choose N}\alpha^N(1-\alpha)^{\tilde N-N}\mathcal{Q}^\text{CO}(N,M,P)},
\end{align}
with $\mathcal{Q}^\text{CO}(\cdot,\cdot,\cdot)$ and $\hat\mu^{N,N}$ given in (\ref{eq:withRngebr2}) and (\ref{eq:RnlebrIC}), respectively.\qed

Finally, Corollary 1 extends the results of Theorems 1, 2 and 4 into the cases with different precoding and fading models.

\textbf{\emph{Corollary 1.}} {Consider the cases with moderate/large number of antennas/users and no or the same path loss for all users. There are mappings between the performance of scheduler-based MIMO networks with different precoders/fading models, in the sense that the sum throughput achieved with different precoding schemes/channel models are the same, as long as the necessary conditions of the CLT are satisfied and one can find appropriate parameter settings such that the considered precoding schemes/channel models lead to the same equivalent Gaussian variables.}
\begin{proof}
{The proof comes from Theorems 1, 2 and 4 where the sum throughput is achieved by taking expectation over the CDF of equivalent Gaussian variables. Thus, from the mathematical perspective, as long as the necessary conditions of the CLT are satisfied and one can find an equivalent Gaussian variable for the sum rate random variable, the only difference between the system performance in different precoding schemes and fading models is in the mean and the variance of the equivalent Gaussian variables, e.g., $(\mu,\sigma^2)$ and $\left(\hat\mu^{N,M},\left(\hat\sigma^{N,M}\right)^2\right)$ in (\ref{eq:eqtheorem1}) and (\ref{eq:eqtheorem2}) for the cases {with no interference cancellation and users' cooperation}, respectively. Thus, if we can set the parameters such that the mean and the variance of the equivalent Gaussian variables of different precoding schemes/fading models are the same, they will lead to the same throughput.}
\end{proof}
{It is important to note that, as opposed to scheduler performance evaluation which may not be feasible via simulations, for every precoding scheme/fading model, if the necessary conditions of the CLT are satisfied,
 %and the equivalent Gaussian variable exists,
 the mean and the variance of the equivalent Gaussian variable can be easily found via simulations or analytically.
%The tightness of the approximations and the performance of the proposed scheduling algorithm are analyzed in Section V.
Finally, note} that the approximation results of Theorem 1, 2 and 4 and Corollary 1 are tight for moderate/high users' activation probabilities such that in each slot the number of data requesting users is high.
\vspace{-0mm}
\subsection{Finite Block-length Analysis}
As a common point, the state-of-the-art results are obtained under the assumption of asymptotically long codewords where the instantaneous achievable rate of a user is given by the Shannon capacity formula $\log(1+\gamma)$ with $\gamma$ standing for the instantaneous received {SINR}. This is an appropriate assumption in the cases with long codewords of length, say, $\gtrsim4000$ channel uses. On the other hand, in many delay-sensitive applications
%, such as the real-time video processing for augmented reality,
the codewords are required to be short (on the order of $\sim 100$ channel uses) and, as a result, $\log(1+\gamma)$ is not an appropriate approximation for the achievable rates \cite{5452208,7134725}.
%Thus, it is interesting to relax the long codeword assumption and investigate the system performance in the presence of finite-length codewords.
As a breakthrough, \cite{5452208} presented bounds/approximations on the achievable rates of finite block-length codes where the maximum achievable information rate for  user $i$ which can be decoded with block error probability no greater than $\varphi_i$ is given by \cite[Theorem. 45]{5452208}
\begin{align}\label{eq:finitemmimo}
r_i\simeq\log(1+\gamma_i)-\sqrt{\frac{1}{L}\left(1-\frac{1}{(1+\gamma_i)^2}\right)}Q^{-1}(\varphi_i).
\end{align}
Here, $L$ is the codeword length
%, $Q^{-1}(\cdot)$ represents the inverse $Q$ function
and $\gamma_i$ denotes the received SINR of user $i.$ Note that the approximation result of (\ref{eq:finitemmimo}) is tight for, say, $L>100$ channel uses on which we concentrate. Moreover, as expected, the achievable rate (\ref{eq:finitemmimo}) increases with the signal length $L$ monotonically and letting $L\to\infty$, (\ref{eq:finitemmimo}) converges to Shannon's capacity formula in the cases with asymptotically long codewords.

Using (\ref{eq:finitemmimo}) and different precoding schemes, one can re-run Algorithm 1 to derive the system performance in the cases with short packets. Also, considering moderate/large number of antennas/users, we can replace the sum of $r_i$'s in (\ref{eq:finitemmimo}) by an equivalent Gaussian variable, with mean and variance as functions of the codeword length $L$ and the constant error probability $\varphi_i=\varphi,\forall i,$ and use the same method as in Theorems 1, 2 and 4 to derive quasi-closed-form expressions for the throughput, and evaluate the effect of codeword length on the system performance.

{Finally, while adaptive power allocation can be easily added into the scheduling approach of Algorithm 1, we concentrated on the cases with equal power allocation. This is because 1) in many practical systems we do not have a lot of dynamic range to adapt the transmit power. Instead, the data is transmitted at a fixed power, and the transmission rates are adapted. Also, 2) the nonadaptive power allocation allows us to derive quasi-closed-form expressions for the maximum achievable rate as given in Section V. Adaptive power allocation is an interesting extension of the paper and is expected to improve the system performance considerably. However, with adaptive power allocation and/or different distances to the users the rate terms in, e.g., (\ref{eq:expectedwithout}) are not IID random variables and the CLT can not be used for approximating the system performance.}

\section{Simulation Results}
Here, we present the simulation results of the proposed GA-based scheduler and verify the accuracy of the derived analytical results. In all figures, except for Fig. 9, we consider Rayleigh fading conditions and set $d_{i,j}=1,\forall i,j$. Performance analysis with different channel models is considered in Fig. 9.
%Also, we have tested the scheduler for different large-scaling fading factors which show the same qualitative conclusions as in the presented figures.
In all figures, except for Fig. 6 which shows an example of the algorithm convergence, we have considered $5\times10^5$ different channel realizations for each point in the simulation curves. Moreover, in all figures, except for Figs. 6 and 11, the algorithm is run for sufficiently large number of iterations until no further performance improvement is observed by increasing the number of iterations. Then, Figs. 6,  11 and Table I study the performance of the proposed scheduler for different numbers of iterations. The results of the algorithm are tested for $K=10$ and $J=5.$ {Also, we have tested the algorithm for the cases where in Step V either one or two users are changed in the queen, and in both cases the algorithm converged to the same results.}
%In the meantime, we have checked the results for other parameter settings of the algorithm as well.
%In Figs. 5b, 7a, 7b, 8 and {11}, we consider zero-forcing precoder. Performance analysis in the cases with {no interference cancellation (resp. users' cooperation)} is considered in Figs. 1a, 2, 3, 6, {9, 10} and {12} (resp. Figs. 1b and 5a). Also, Fig. 4 compares the performance of different precoding schemes.
\vspace{-0mm}
\subsection{On the Accuracy of the Analytical Results}
As discussed in Example 1, with typical numbers of antennas/users, analyzing the exhaustive search-based optimal scheduler implies very large number {of scheduling rule checkings}. As a result, it is not feasible to compare the analytical approximations with the exhaustive search based approach, unless for small networks (see Fig. 4). For this reason, as an appropriate metric to measure the difference between PDFs, we first analyze the Kullback-Leibler divergence (KLD) \cite[p. 22]{4444444444} of the PDF of the random variable ${Z_n^{N,M}}$ (resp. ${W_n^{N,M}}$) in (\ref{eq:Zdef}) (resp. (\ref{eq:Zdefout})) and its corresponding CLT-based Gaussian approximation $\mathcal{Z}$ (resp. $\mathcal{Y}$) in the cases {with no  interference cancellation (resp. users' cooperation)}. {In mathematical statistics, the KLD (also called relative entropy) is a measure of how one PDF diverges from another PDF. Particularly, a large value of KLD indicates different characteristics of two random variables, while the KLD converges to zero as the PDFs of two random variables become identical.} Note that the CLT-based approximations of ${Z_n^{N,M}}$ and ${W_n^{N,M}}$ are the only approximations that we have applied in Theorems 1 and 4. Thus, a low value of KLD well {proves} the accuracy of the approximations. Then, we compare the analytical results with those achieved by GA-based scheduler running for a large number of iterations which tightly mimic the exhaustive search approach.

Figures 1a and 1b demonstrate the KLD of the empirical PDF of the achievable rates and their CLT-based Gaussian approximations for the cases {with no interference cancellation and users' cooperation}, respectively. Here, KLD is defined as \cite[p. 22]{4444444444}
\begin{align}\label{eq:eqKLDno}
D\left(f_{Z_n^{N,M}},f_\mathcal{Z}\right)=\int_0^\infty{f_{Z_n^{N,M}(x)}\log\left(\frac{f_{Z_n^{N,M}}(x)}{f_\mathcal{Z}(x)}\right)\text{d}x},
\end{align}
and
\begin{align}\label{eq:eqKLDno}
D\left(f_{W_n^{N,M}},f_\mathcal{Y}\right)=\int_0^\infty{f_{W_n^{N,M}}(x)\log\left(\frac{f_{W_n^{N,M}}(x)}{f_\mathcal{Y}(x)}\right)\text{d}x},
\end{align}
in the cases {with no  interference cancellation and users' cooperation}, respectively. Also, Fig. 1b evaluates the KLD in the cases where the mean and variance of the equivalent Gaussian distribution are given by the low-SNR approximation (\ref{eq:tarokhapp2}). Then, Fig. 2 compares the empirical and the approximated PDFs of the achievable sum rates $\sum_{i=1}^N{\log\left(1+\frac{\frac{P}{M} g_{i,i}}{\frac{P}{M}\sum_{j=1,j\ne i}^{N}{ g_{i,j}}+1}\right)}$ in (\ref{eq:expectedwithout2}) for the cases with no interference cancellation.

\begin{figure}
\centering
  \includegraphics[width=0.95\columnwidth]{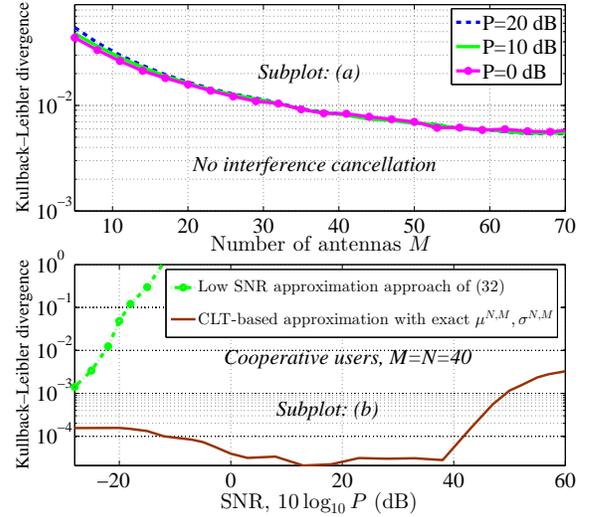}\\\vspace{-2mm}
\caption{Kullback-Leibler divergence between the empirical and the analytically approximated PDFs of the {users' sum} rates in the cases with (a): no interference cancellation, and (b): {users' cooperation}. The results are presented for the cases with $M$ transmit antennas and $N=M$ users.}\vspace{-2mm}\label{figure111}
\end{figure}

\begin{figure}\label{figMMIMOeffectofPA}
\centering
  \includegraphics[width=0.95\columnwidth]{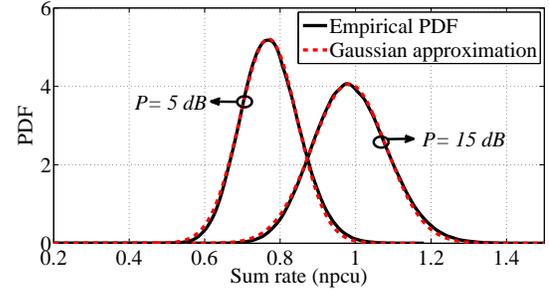}\\\vspace{-2mm}
\caption{Empirical and approximated PDFs of the {sum} rate, no interference cancellation, $M=N=60.$}\vspace{-2mm}\label{figure111}
\end{figure}

As it can be seen in Figs. 1-2, for a broad range of SNRs/number of antennas, the CLT-based approximation and the approximation approach of (\ref{eq:tarokhapp2}) result in very low values of KLD, and the achievable rates can be approximated by equivalent Gaussian variables with high accuracy. {Also, compared to the cases approximating the mean and the variance of the equivalent Gaussian variable, the tightness of the CLT-based approximation increases when the mean and the variance are found by simulations (Fig. 1b).} Thus, the CLT-based approach and (\ref{eq:tarokhapp2}) provide  tight approximations for the ultimate performance of the optimal scheduler. Finally, in harmony with intuitions, the KLD decreases with the number of antennas, i.e., the tightness of the CLT-based approximation increases with the network size.

\begin{figure}
\centering
  \includegraphics[width=0.95\columnwidth]{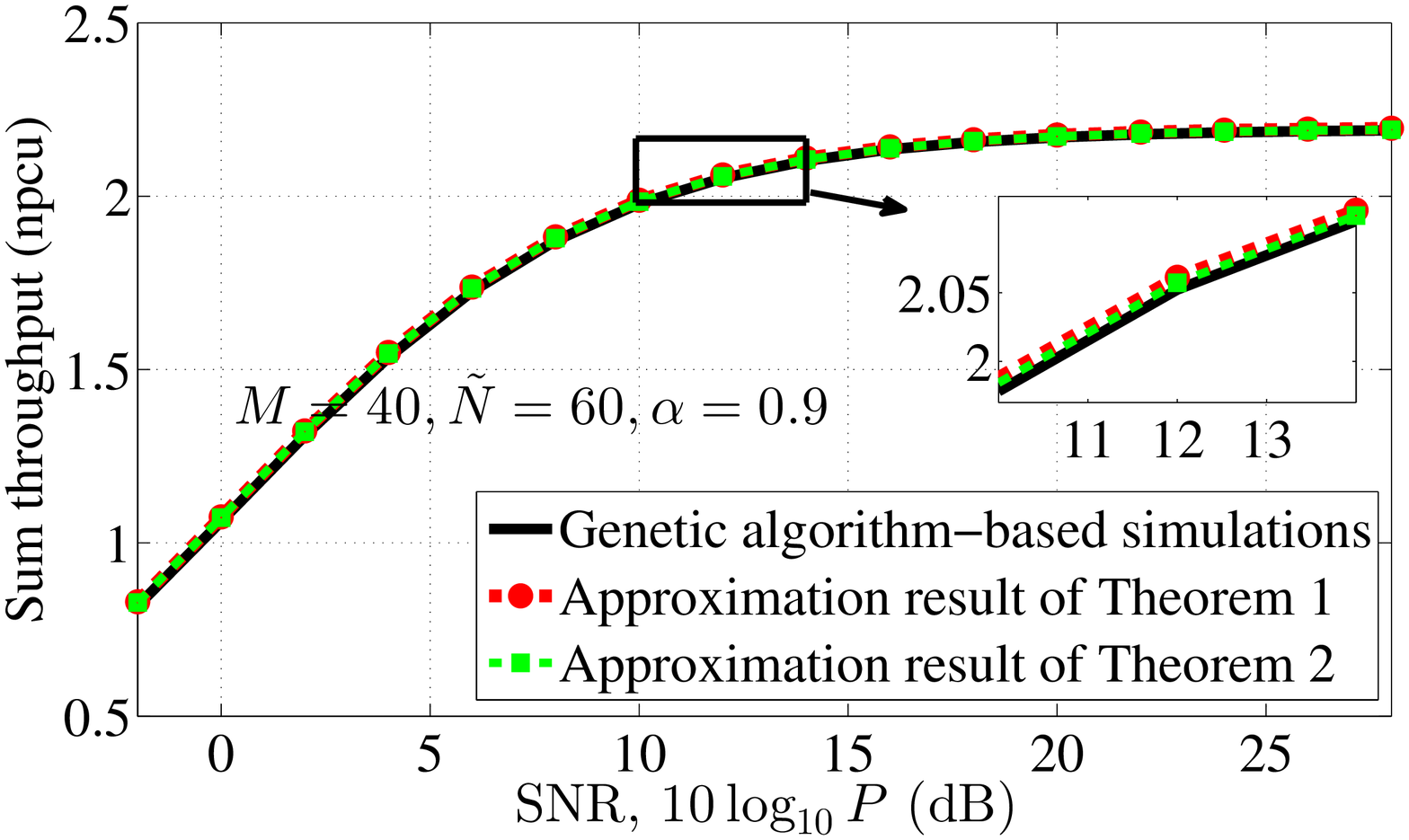}\\\vspace{-2mm}
\caption{On the tightness of Theorems 1-2, no interference cancellation, $\alpha=0.9, M=40$, and $\tilde N=60.$}\vspace{-2mm}\label{figure111}
\end{figure}

\begin{figure}
\centering
  \includegraphics[width=0.95\columnwidth]{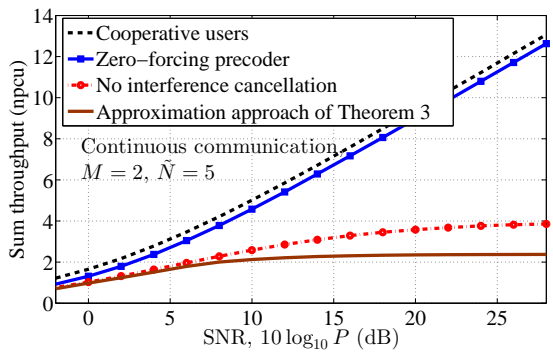}\\\vspace{-2mm}
\caption{Performance analysis in small networks, continuous communications,  $\alpha=1, M=2$, and $\tilde N=5.$}\vspace{-2mm}\label{figure111}
\end{figure}

\begin{figure}
\centering
  \includegraphics[width=0.95\columnwidth]{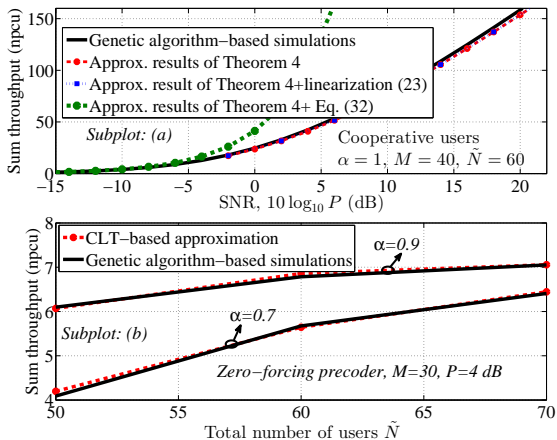}\\\vspace{-2mm}
\caption{On the tightness of the analytical results. (a) Sum throughput for the cases with {users' cooperation}, $\alpha=1, M=40, \tilde N=60.$ (b): The simulation and the analytical performance analysis of the zero-forcing based precoder, $M=30, P=4$ dB.}\vspace{-2mm}\label{figure111}
\end{figure}

Setting $M=40, \tilde N=60, \alpha=0.9,$ Fig. 3 verifies the tightness of the approximation schemes of Theorems 1-2 for the cases with no interference cancellation. Then, Fig. 4 studies the achievable sum throughput in the cases with no interference cancellation, zero-forcing precoder and {users' cooperation}, and compares the results with those achieved via the approximation approach of Theorem 3. Here, the results are presented for continuous communications ($\alpha=1$), $M=2$ and $\tilde N=5.$ Also, for different precoding schemes, the simulation results have been derived by both exhaustive search, which is feasible for the considered parameter setting of Fig. 4, and GA-based approach, and they lead to the same system throughput.

The tightness of the approximation scheme of Theorem 4 is evaluated in Fig. 5a, where we study the sum throughput in the cases with $M=40, \tilde N=60,$ continuous communications and {users' cooperation}. Also, the figure studies the accuracy of the CLT-based approximation in the cases where the one-dimensional integration of (\ref{eq:eqtheorem2}) is approximated by the same linearization techniques as in (\ref{eq:eqlinearapprox}) as well as the cases where the mean and the variance of the equivalent Gaussian variable are approximated by (\ref{eq:tarokhapp2}). Finally, to validate Corollary 1, Fig. 5b compares the achievable sum throughput of the zero-forcing based scheduler with the cases where the same CLT-based approximation technique as in Theorems 1, 4 is used to derive the network throughput analytically. That is, in Fig. 5b the analytical results are obtained by finding the mean and variance of the achievable rate term in (\ref{eq:expectedwithout2}) {through simulations} for the cases with zero-forcing precoder, and replacing them in the throughput expression (\ref{eq:eqtheorem2}). Here, the results are presented for the cases with $M=30, P=4$ dB, perfect CSI available at the transmitter and different total numbers of users $\tilde N.$

As it is observed in Figs. 1-5, the analytical results of Theorems 1, 2 and 4 mimic the exact results with very high accuracy. Also, the linearization technique of (\ref{eq:eqlinearapprox}) can effectively be applied to derive a tight quasi-closed-form approximation for the achievable throughput of the optimal scheduler in the cases with no  interference cancellation and users' cooperation. %Moreover, Theorem 3 properly lower-bounds the network throughput and the tightness increases at low SNRs/number of antennas (Fig. 4).
{Moreover, Theorem 3 properly lower-bounds the network throughput and leads to a tight bound at low SNRs/number of antennas (Fig. 4). At high SNRs, however, the tightness of the lower bound of Theorem 3 decreases. This is intuitive because the difference between the CDFs $F_{v_i^{N,M}}$ and $F_{u_i^{N,M}}$ increases with the SNR (see Theorem 3)}. At low/moderate SNRs, (\ref{eq:tarokhapp2}) provides tight approximations for the mean and variance of the equivalent Gaussian variable and the corresponding approximation matches the exact values derived via simulations with high accuracy (Fig. 5a). Also, in harmony with Corollary 1, the same CLT-based approximation as in Theorems 1 and 4 can be applied for the cases with, e.g., zero-forcing precoder, as long as the mean and variance of the equivalent Gaussian variables are given (see Fig. 5b and Corollary 1). Finally, the tightness of the CLT-based approximation and, consequently, the approximation schemes of Theorems 1, 2, 4, increases with the number of antennas.
%This is because the tightness of the CLT-based approximations increases with $M$.

In summary, the results of Figs. 1-5 are interesting from two perspectives:
\begin{itemize}
  \item[1)] According to the figures, the approximation schemes of Theorems 1, 2 and 4 (resp. Theorem 3) provide effective tools for the analytical investigation of the optimal scheduler in the cases with large (resp. small) number of antennas/users. Also, the derived quasi-closed-form expressions can be utilized as a {benchmark} to evaluate the efficiency of different sub-optimal schedulers, e.g., \cite{1705935,5360758,1492685,m06094252,m07127500,m06884146,m06831724,06095627,m07037316,m06941317,06850064,05456039,05336871,01312486,01664083,04635028,04668538,04570215,04299613,05374079,05871799,06504540,06878495,06740115,06725595,05999732,06810624,04784356}.
  \item[2)] While the exhaustive search-based optimal scheduling is not feasible in the cases with moderate/large number of antennas/users, the proposed GA-based scheduler results in (almost) the same throughput as in the derived quasi-closed-form expressions for the optimal scheduler (Figs. 3, 5). That is, while Section V derives quasi-closed-form expressions for the ultimate performance of the optimal schedulers, Section IV develops an effective approach to reach the ultimate system performance with few iterations.
\end{itemize}

Finally, as a side result, Fig. 4 indicates the efficiency of the zero-forcing precoder where the gap between the throughput of the cases with zero-forcing precoder and {users' cooperation} is relatively small for a broad range of SNRs. Also, while interference cancellation improves the system performance significantly as the transmission power/number of antennas increases, it leads to marginal throughput increment at low SNRs.
%Finally, while the sum throughput increases with the total number of
\vspace{-0mm}
\subsection{On the Performance of the GA-based Scheduler}
In Figs. 6-12 and Table I, we study the performance of the proposed GA-based scheduler and evaluate the effect of different parameters on the system throughput. The simulation results are presented in different parts as follows.

\begin{figure}
\centering
  \includegraphics[width=0.95\columnwidth]{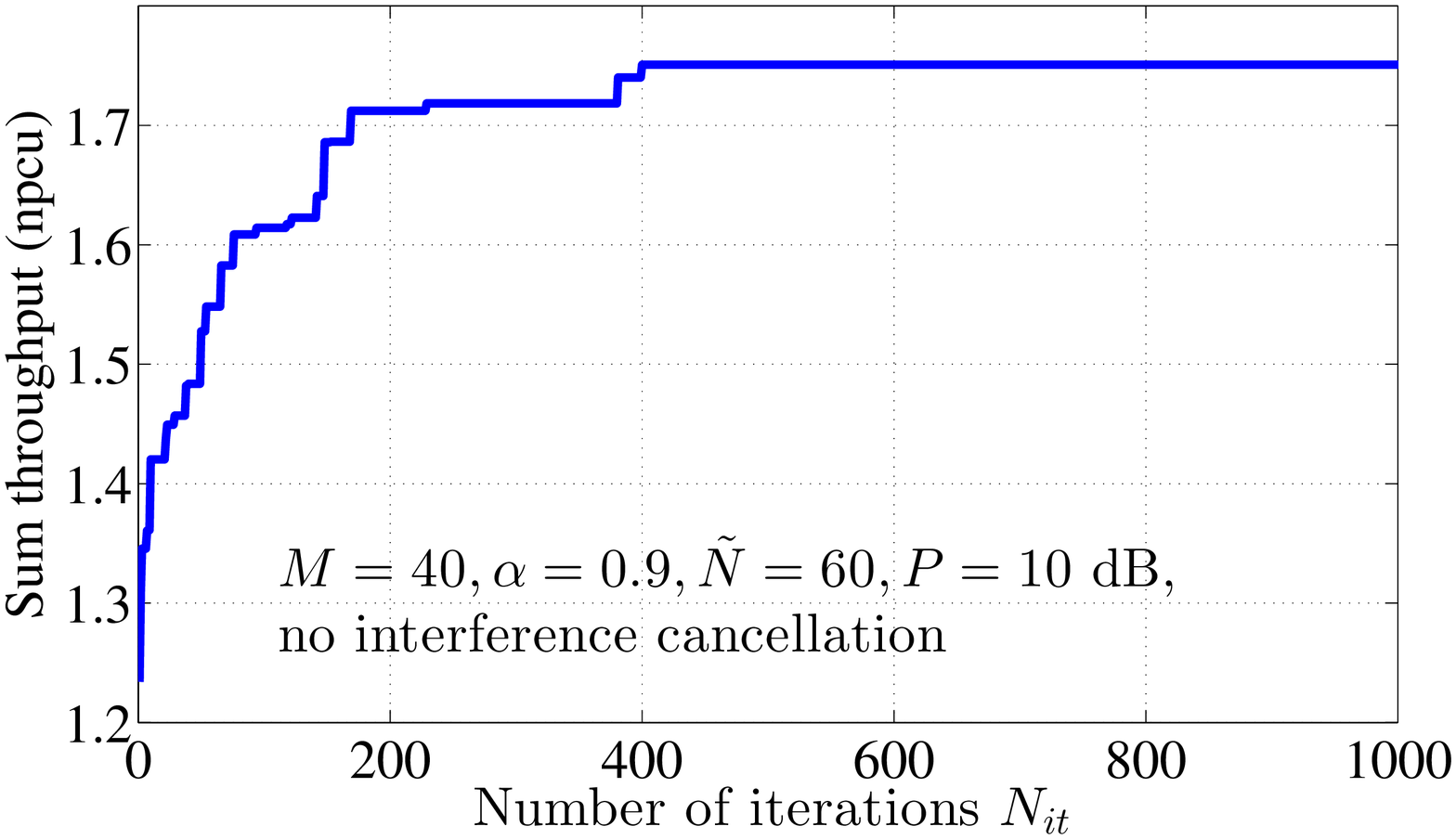}\\\vspace{-2mm}
\caption{An example of the convergence process of the proposed algorithm. {The results are presented for a single channel realization, bursty} communications, no interference cancellation, $\tilde N=60,$ $M=40, \alpha=0.9,$ and  $P=10\text{ dB}$.}\vspace{-2mm}\label{figure111}
\end{figure}

\begin{table}\caption{Average numbers of iterations required for the convergence of the algorithm, bursty communications, $\alpha=0.9$. The results of the GA-based scheme are obtained by averaging over $5\times10^5$ different random channel realizations. }\vspace{-2mm}
\center
\begin{tabular}{cc|c|c|c|c|}
\cline{3-5}
& & \multicolumn{3}{ c| }{Number of transmit antennas} \\ \cline{3-5}
& & $M=20$ & $M=30$ & $M=40$ \\ \cline{1-5}
\multicolumn{1}{ |c  }{\multirow{2}{*}{$\tilde N=60$} } &
\multicolumn{1}{ |c| }{GA-based} & $472$ & $487$ & $451$       \\ \cline{2-5}
\multicolumn{1}{ |c  }{}                        &
\multicolumn{1}{ |c| }{Exhaustive search-based} & $5\times 10^{14}$ & $5\times 10^{15}$ & $6\times 10^{13}$       \\ \cline{1-5}
\multicolumn{1}{ |c  }{\multirow{2}{*}{$\tilde N=80$} } &
\multicolumn{1}{ |c| }{GA-based} & $668$ & $746$ & $767$    \\ \cline{2-5}
\multicolumn{1}{ |c  }{}                        &
\multicolumn{1}{ |c| }{Exhaustive search-based} & $4\times 10^{17}$ & $3\times 10^{20}$ & $1\times 10^{21}$    \\ \cline{1-5}
\end{tabular}\vspace{-2mm}
\end{table}

%\begin{figure}
%\centering
%  % Requires \usepackage{graphicx}
%  \includegraphics[width=0.5\columnwidth]{bFigMMIMOLOSkn.eps}\\\vspace{-6mm}
%\caption{Sum throughput for different line-of-sight channel factors $\nu$, bursty communications, zero-forcing precoder, $\alpha=0.9, M=30$, and $\tilde N=60.$}\vspace{-11mm}\label{figure111}
%\end{figure}

\emph{On the performance of the proposed algorithm:} {Considering a single channel realization, }Fig. 6 shows an example for the convergence of the proposed GA-based scheduler in the cases with $M=40, \tilde N=60, K=10, J=5, \alpha=0.9,$ and  $P=10 \text{ dB}$. Then, setting $K=10,$ Table I shows the average number of iterations that are required in the proposed GA-based scheduler to achieve (almost) the same throughput as in the optimal exhaustive search-based scheduler.
%Note that the exhaustive search-based results can be achieved by letting for a large number of iterations in Algorithm 1.
Also, the table compares the required number of iterations of the proposed algorithm with those in exhaustive search which are found by (\ref{eq:eqexpectedcheckings}). The results of Table I are presented for the users' data request probabilities $\alpha=0.9.$ {Also, the algorithm is stopped if no improvement is observed after a number of iterations ($500$ iterations in the simulations of Table I).}

From Fig. 6, we observe {that the} system performance improves with the number of iterations monotonically. However, the developed GA-based method leads to (almost) the same performance as the exhaustive search-based scheduler with very limited number of iterations (Fig. 6, Table I). For example, with the parameter settings of Fig. 6, our GA-based scheduler reaches more than $95\%$ of the maximum achievable throughput with less than $200$ iterations (note that with the parameter settings of Fig. 6 the exhaustive search implies $6\times 10^{13}$ different parameter checkings). Thus, the proposed algorithm can be effectively applied for user scheduling in the cases with large-but-finite number of antennas/users. Finally, the algorithm converges in a ladder fashion. This is because the system performance is not necessarily improved in each iteration, and it may fall into a local optimum in some iterations. However, because of Steps V-VII of the algorithm, it can always escape a local minima and reach the global optimum, if sufficiently large number of iterations are considered (Fig. 6).

\begin{figure}
\centering
  \includegraphics[width=0.95\columnwidth]{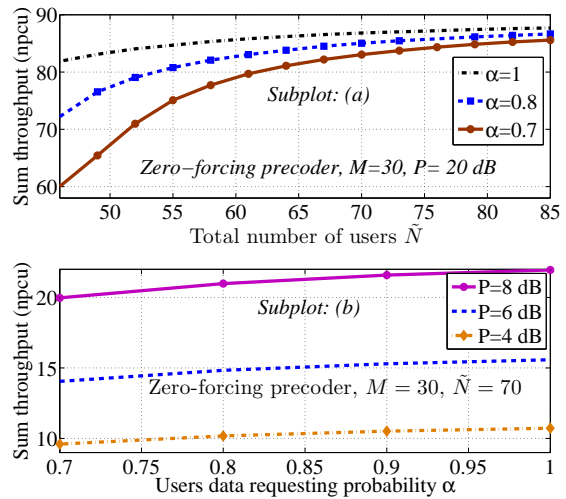}\\\vspace{-2mm}
\caption{Sum throughput vs (a) the total number of users $\tilde N$, and (b) users' data request probability $\alpha.$ The results are presented for zero-forcing precoder and $M=30.$ In Subplots (a) and (b), we set $P=20$ dB, and  $\tilde N=70,$ respectively.}\vspace{-2mm}\label{figure111}
\end{figure}

\emph{On the performance of the zero-forcing based precoder:} Figure 7a studies the sum throughput of the scheduler-based network for different maximum numbers of users $\tilde N$. Here, the results are presented for the cases with $M=30, P=20$ dB, and zero-forcing precoder. Then, considering bursty communications, $M=30$ and $\tilde N=70,$ Fig. 7b shows the network sum throughput for different users' data request probabilities $\alpha.$ Finally, setting $M=30, \tilde N=70,$ Fig. 8 demonstrates the system performance for different SNRs. As seen in Figs. 7a and 7b, the performance of the zero-forcing based scheduler is improved by increasing the number of users and the users' data request probabilities. This is {intuitive} because with a large number of  users asking for data the scheduler has better chance to select the users with high channel quality. However, the relative effect of the number of users and their data request probabilities decreases as $\tilde N$ and $\alpha$ increase (also, see Fig. 5b). Also, the sensitivity of the throughput to the users' data request probability increases as transmission power increases/maximum number of users $\tilde N$ decreases (Figs. 7-8).

\begin{figure}
\centering
  \includegraphics[width=0.95\columnwidth]{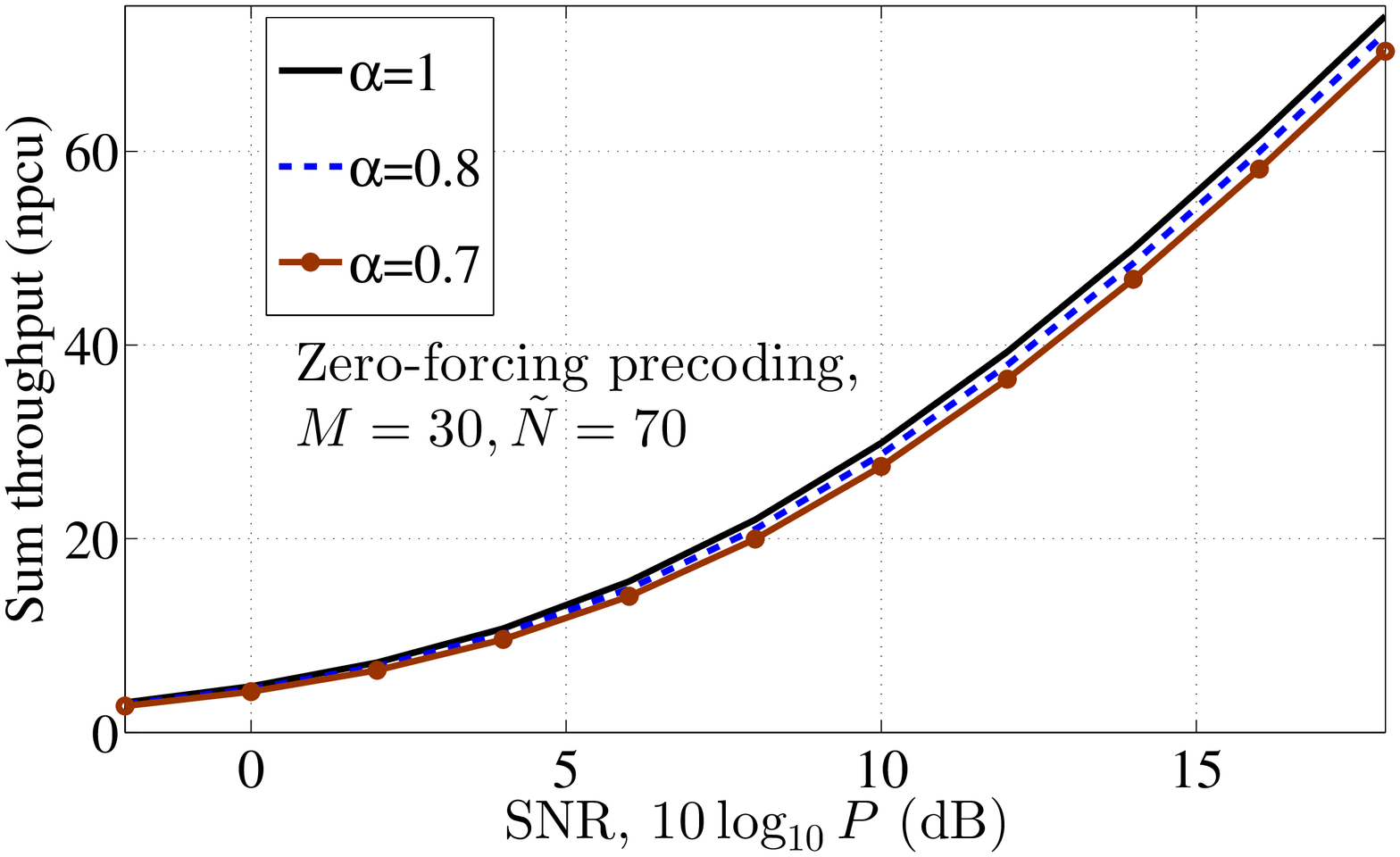}\\\vspace{-2mm}
\caption{Sum throughput vs the SNR, zero-forcing precoder, $M=30$, and $\tilde N=70.$}\vspace{-2mm}\label{figure111}
\end{figure}

\emph{Performance analysis in different channel models/number of scheduled users:}
{While we presented the analytical and simulation results for the cases with no path loss and Rayleigh fading, the GA-based scheduler is well applicable for different path loss/channel models. Also, in Sections IV-V, we presented the results for the cases where, if at least $M$ users ask for data, then $M$ users will be scheduled. This is motivated by serving as many as possible users and also because of the implementation complexity of the scheduling algorithm. However, it is straightforward to modify Algorithm 1 such that the number of served users are also optimized. For instance, considering different values of path loss exponent in (\ref{eq:channelmodel}), Fig. 9 demonstrates the system throughput for the cases optimizing the number of served users and compares the results with those obtained by always serving $M$ users out of $N\ge M$ data requesting users. Here, the results are presented for the cases with $M=7, \tilde N=20,$ users randomly dropped in a circle with radius $50$ m, continuous communications and no interference cancellation.}

\begin{figure}
\centering
  \includegraphics[width=0.95\columnwidth]{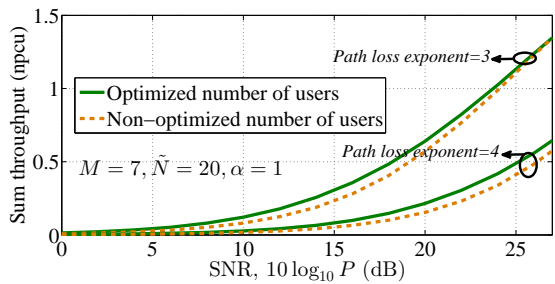}\\\vspace{-2mm}
\caption{Sum throughput for different path loss models in (\ref{eq:channelmodel}) and optimal number of served users. Continuous communications, no interference cancellation,  $M=7$, and $\tilde N=20.$}\vspace{-2mm}\label{figure111}
\end{figure}

{As seen in Fig. 9, the proposed algorithm can be effectively applied for different channel models. Also, while the system performance is improved by optimizing the number of served users in each time slot, the performance improvement, compared to the cases always serving $M$ users out of $N\ge M$ data requesting users, is negligible for a broad range of SNRs and path loss exponents. Also, the relative performance gain of optimizing the number of served users is observed at higher SNRs, as the path loss exponent increases. Finally, note that the CLT-based approximation results of Theorems 1, 2 and 4 can be extended to the cases with optimized number of served users.}

\emph{On the effect of imperfect power amplifiers:} In Sections IV-V, we considered perfect hardware. However, as the number of antennas increases, the hardware inefficiency, especially the power amplifiers' inefficiency may affect the system performance remarkably. Interestingly, the proposed algorithm can be adopted to take different hardware impairments into account. For this reason, Fig. {10} studies the performance of the scheduler-based network in the cases with imperfect power amplifiers and no interference cancellation. Particularly, we consider the state-of-the-art
power amplifiers' efficiency model where the output power at the $m$-th antenna is given by \cite{phdthesisBjornemo,4160747}, \cite[Eq. (3)]{6725577}, \cite[Eq. (1)]{7104158}
\vspace{-0mm}
\begin{align}\label{eq:ampmodeldaniel}
&\frac{P_m}{P_m^\text{cons}}=\epsilon\left(\frac{P_m}{P_m^\text{max}}\right)^\vartheta\,\Rightarrow  P_m=\left({\frac{\epsilon P_m^\text{cons}}{(P_m^\text{max})^\vartheta}}\right)^\frac{1}{1-\vartheta}, m=1,\ldots,M.
\end{align}
Here, $P_m, P_m^\text{max}$ and $P_m^\text{cons}$ are the output, the maximum output, and the consumed power of the $m$-th antenna, respectively, $\epsilon\in [0,1]$ denotes the maximum power efficiency achieved at $P_m=P_m^\text{max},\forall m,$ and $\vartheta$ is a parameter that, depending on the power amplifier classes, varies between $[0,1]$. Also, the total consumed power at the transmitter is given by $P^\text{cons}=MP_m^\text{cons}.$ The results of Fig. 10 are given for $M=40, \tilde N=60, \vartheta=0.5,$ $P_m^\text{max}=25 $ dB, $\forall m,$ and different total consumed powers. As seen, the inefficiency of the power amplifiers affects the throughput significantly. For instance, with the parameter settings of Fig. 10 and the throughput 1 npcu, decreasing the power amplifiers' efficiency from $45\%$ to $30\%$ leads to $2$ dB loss in power efficiency. Also, with the total consumed {power $56$ dBm,} the throughput reduces by $50\%$ when the power amplifiers' efficiency decreases from $75\%$ to $25\%.$ Thus, the power amplifiers' efficiency should be carefully taken into account in the network design.

\begin{figure}\label{figMMIMOeffectofPA}
\centering
  \includegraphics[width=0.95\columnwidth]{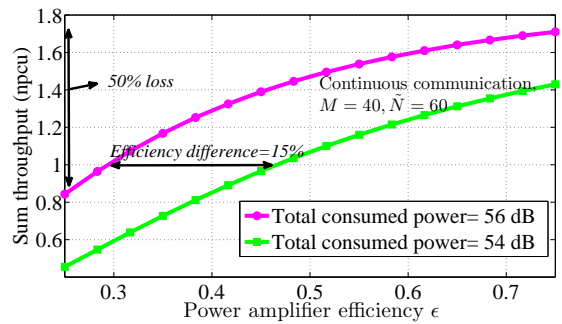}\\\vspace{-2mm}
\caption{Sum throughput versus the power amplifiers' efficiency, continuous communications, no precoder, $M=40$, and $\tilde N=60.$}\vspace{-2mm}\label{figure111}
\end{figure}

\emph{On the cost of scheduling:} As the network size increases, the scheduling delay may reduce the end-to-end throughput. To analyze the cost of the scheduling, we consider the end-to-end throughput defined as
\vspace{-0mm}
\begin{align}\label{eq:eqendtoendeta}
\eta^\text{end-to-end}(k)=\left(1-{\xi}k\right)\sum_{N=1}^{\tilde N}{\Pr(N)\mathbb{E}\{R^k(N)\}}.
\end{align}
Here, $\mathbb{E}\{R^k(N)\}$ represents the expected achievable rate with $N$ data requesting users and the scheduling rule in the $k$-th iteration of the algorithm. Also, $\xi$ denotes the normalized delay for each iteration of the algorithm (normalized by the total packet length), i.e., the delay cost for each iteration of the algorithm.
\begin{figure}\label{figMMIMOeffectofPA}
\centering
  \includegraphics[width=0.95\columnwidth]{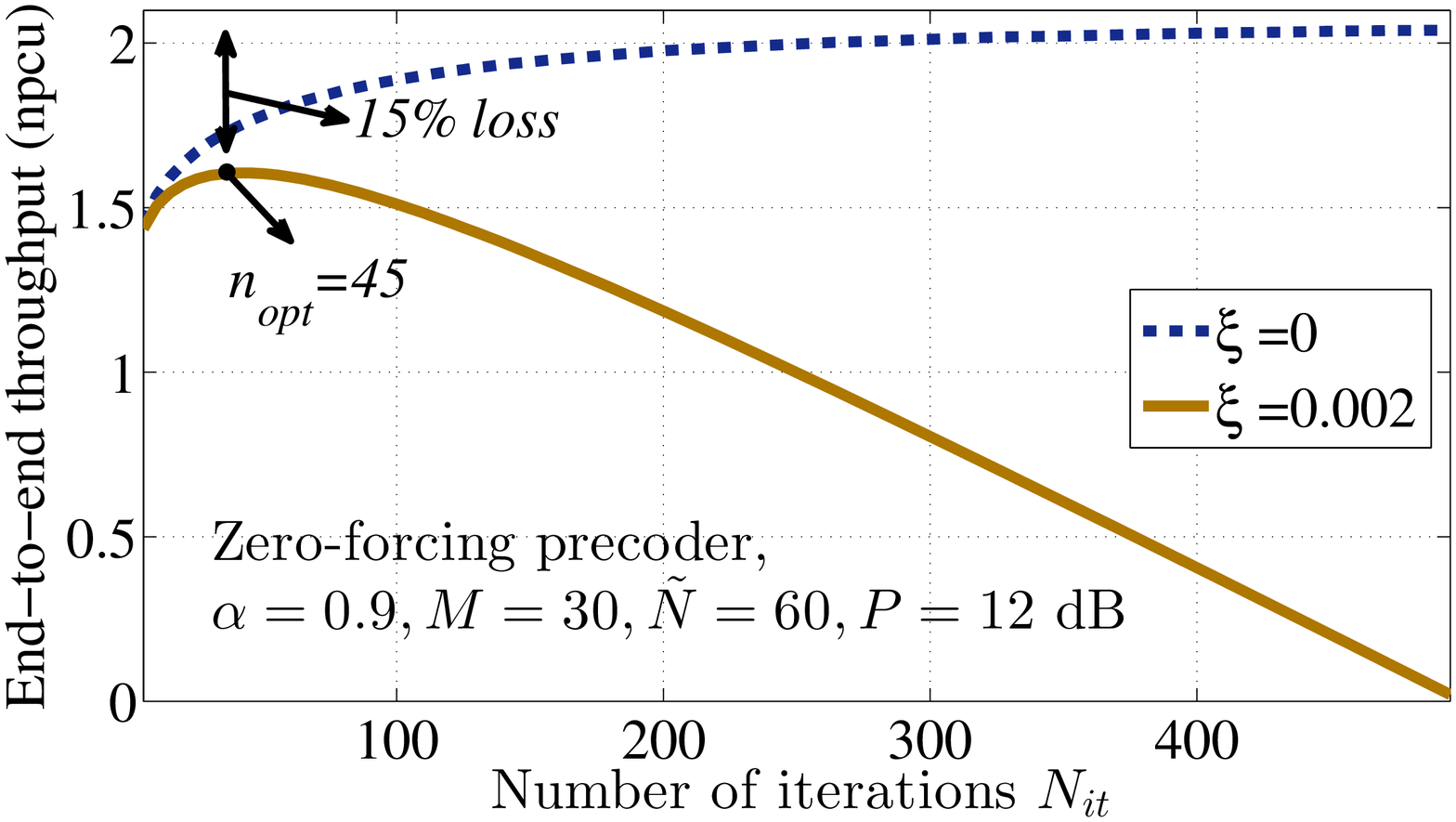}\\\vspace{-2mm}
\caption{End-to-end throughput in the cases with different iteration costs of the algorithm, zero-forcing precoder, $\alpha=0.9, M=30, \tilde N=60, P=12$ dB.}\vspace{-2mm}\label{figure111}
\end{figure}

Considering zero-forcing precoder, $\alpha=0.9, M=30, \tilde N=60,$ and $P=12$ dB, Fig. {11} studies the network end-to-end throughput versus the maximum number of iterations $N_\text{it}$. Note that, in contrast to Fig. 6 representing the algorithm performance for a single channel realization, the results of Fig. {11} are obtained by averaging the system performance over $5\times 10^5$ different channel realizations. As a result, the ladder-shape convergence is smoothed out in Fig. {11}.

As opposed to the delay-insensitive scenario, i.e., $\xi=0,$ where the system performance improves monotonically with $N_\text{it}$ until it converges to the exhaustive search-based results, the end-to-end throughput does not necessarily improve with the  number of iterations if the delay cost of the algorithm is taken into account (see Fig. {11}, the case with $\xi=0.002$). That is, considering the algorithm running delay, there is a trade-off between finding the optimal scheduling and reducing the data transmission time slot. Thus, the highest throughput may be achieved by few iterations, i.e., a rough estimation of the optimal scheduler, and there is an optimal number of iterations maximizing the throughput.
%For instance, with the parameter settings of Fig. 10b and $\xi=0.002$, the maximum end-to-end throughput is achieved by selecting a sub-optimal scheduling after $n_\text{opt}=45$ iterations with $85\%$ of the performance of the optimal scheduler, and leaving the rest of the packet for data transmission.
{For instance, with the parameter settings of Fig. 11 and $\xi=0.002$, the maximum end-to-end throughput is achieved by selecting a sub-optimal scheduling after $n_\text{opt}=45$ iterations, and leaving the rest of the packet for data transmission. However, as seen in Fig. 11, with 45 iterations the algorithm finds queens which lead to $85\%$ throughput of the optimal exhaustive-search based results, i.e., $15\%$ throughput loss. Then, using (\ref{eq:eqendtoendeta}) with $\xi=0.002$ and $k=45,$ the end-to-end throughput would be $77\%$ of the throughput of the optimal scheduler with no cost. }Finally, note that, while we concentrate on temporally-independent quasi-static conditions, in practice there may be considerable correlation between the channel realizations in successive fading blocks. In that case, the queen of the previous fading block can be an appropriate initial guess for the optimal scheduling rule of the next block and, as a result, the required number of  iterations may decrease significantly.
%the end-to-end throughput converges to zero at $k=\frac{1}{\xi}$ (see Fig. 10b and (\ref{eq:eqendtoendeta})).

\emph{Finite block-length analysis:} Considering $M=30, \tilde N=60, P=14, 16$ dB, and bursty communications with $\alpha=0.9$, Fig. {12} analyzes the system throughput in the cases with short packets. Here, the results are obtained for different codewords lengths $L$ and {block} error probabilities $\varphi$ in (\ref{eq:finitemmimo}). {Also, with finite block-length codes and in harmony with, e.g., \cite{5452208,8003368}, we define the throughput as $\eta=\sum_{N=1}^{\tilde N}{\Pr(N)\mathbb{E}\{r(N)\}}$ with $r(N)=\sum_{i=1}^{N}r_i$ and $r_i$ given in (\ref{eq:finitemmimo}).} As seen, the proposed algorithm is well applicable in the cases with different codewords lengths. Also, in harmony with intuitions, the achievable throughput increases as the required {block} error probability tends towards one. Particularly, at low error probabilities the throughput increases (almost) logarithmically with the {block} error probability. On the other hand, the throughput decreases significantly in the cases with strict {block} error probability requirements, i.e., small $\varphi$'s. Finally, the system throughput is sensitive to the length of short packets while its sensitivity to the packets length decreases for long packets.

\begin{figure}\label{figMMIMOeffectofPA}
\centering
  \includegraphics[width=0.95\columnwidth]{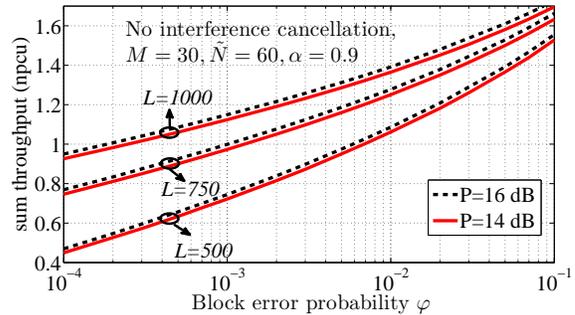}\\\vspace{-2mm}
\caption{Finite block-length analysis of the scheduler-based network, no interference cancellation, $\alpha=0.9, M=30, \tilde N=60, P=14, 16$ dB.}\vspace{-2mm}\label{figure111}
\end{figure}

\vspace{-0mm}
\section{Conclusion}
This paper studied the performance of the scheduler-based MIMO  networks in the cases with large-but-finite number of antennas/users. Considering different communication/channel models and precoding schemes, we presented quasi-closed-form expressions for the ultimate performance of the optimal scheduler. Also, we presented mappings between the performance of different precoding schemes, in the sense that with appropriate parameter settings the same throughput is achieved in these setups. Finally, we proposed an efficient scheduling approach based on GAs and evaluated the system performance for different channel conditions. As demonstrated, the proposed scheduler can reach (almost) the same performance as the optimal scheduler with few iterations. Also, while the network sum throughput is not sensitive to the precoding scheme at low SNRs, significant performance improvement is achieved by, e.g., zero-forcing precoding as the number of antennas/transmission power increases. Finally, the scheduling running delay, the hardware impairments and the length of short packets affect the performance of the large MIMO networks remarkably and should be carefully considered in the network design.
\appendices
\vspace{-0mm}

\vspace{-0mm}

\bibliographystyle{IEEEtran} %lic.bst is the style file
\bibliography{masterMMIMO}

\begin{thebibliography}{10}
\providecommand{\url}[1]{#1}
\csname url@samestyle\endcsname
\providecommand{\newblock}{\relax}
\providecommand{\bibinfo}[2]{#2}
\providecommand{\BIBentrySTDinterwordspacing}{\spaceskip=0pt\relax}
\providecommand{\BIBentryALTinterwordstretchfactor}{4}
\providecommand{\BIBentryALTinterwordspacing}{\spaceskip=\fontdimen2\font plus
\BIBentryALTinterwordstretchfactor\fontdimen3\font minus
  \fontdimen4\font\relax}
\providecommand{\BIBforeignlanguage}[2]{{%
\expandafter\ifx\csname l@#1\endcsname\relax
\typeout{** WARNING: IEEEtran.bst: No hyphenation pattern has been}%
\typeout{** loaded for the language `#1'. Using the pattern for}%
\typeout{** the default language instead.}%
\else
\language=\csname l@#1\endcsname
\fi
#2}}
\providecommand{\BIBdecl}{\relax}
\BIBdecl

\bibitem{8377145}
B.~Makki, T.~Svensson, and M.~Alouini, ``Throughput analysis of
  large-but-finite {MIMO} networks using schedulers,'' in \emph{IEEE
  WCNC'2018}, Bacelona, Spain, April 2018, pp. 1--6.

\bibitem{1705935}
V.~K.~N. Lau, ``Coverage-optimized downlink scheduling design for wireless
  systems with multiple antennas,'' \emph{IEEE Trans. Wireless Commun.},
  vol.~5, no.~10, pp. 2734--2741, Oct. 2006.

\bibitem{5360758}
R.~C. Elliott, S.~Sigdel, W.~A. Krzymien, M.~Al-Shalash, and A.~C.~K. Soong,
  ``Genetic and greedy user scheduling for multiuser {MIMO} systems with
  successive zero-forcing,'' in \emph{IEEE Globecom Workshops}, Hawaii, USA,
  Nov. 2009, pp. 1--6.

\bibitem{1492685}
V.~K.~N. Lau, ``Optimal downlink space-time scheduling design with convex
  utility functions-multiple-antenna systems with orthogonal spatial
  multiplexing,'' \emph{IEEE Trans. Veh. Technol.}, vol.~54, no.~4, pp.
  1322--1333, July 2005.

\bibitem{04668538}
R.~C. Elliott and W.~A. Krzymien, ``Downlink scheduling via genetic algorithms
  for multiuser single-carrier and multicarrier {MIMO} systems with dirty paper
  coding,'' \emph{IEEE Trans. Veh. Technol.}, vol.~58, no.~7, pp. 3247--3262,
  Sept. 2009.

\bibitem{m06094252}
H.~Huh, S.~H. Moon, Y.~T. Kim, I.~Lee, and G.~Caire, ``Multi-cell {MIMO}
  downlink with cell cooperation and fair scheduling: A large-system limit
  analysis,'' \emph{IEEE Trans. Inf. Theory}, vol.~57, no.~12, pp. 7771--7786,
  Dec. 2011.

\bibitem{m07127500}
M.~Benmimoune, E.~Driouch, W.~Ajib, and D.~Massicotte, ``Joint transmit antenna
  selection and user scheduling for massive {MIMO} systems,'' in \emph{IEEE
  WCNC'2015}, New Orleans, LA, USA, March 2015, pp. 381--386.

\bibitem{m06884146}
Y.~Xu, G.~Yue, N.~Prasad, S.~Rangarajan, and S.~Mao, ``User grouping and
  scheduling for large scale {MIMO} systems with two-stage precoding,'' in
  \emph{IEEE ICC'2014}, Sydney, Australia, June 2014, pp. 5197--5202.

\bibitem{m06831724}
Q.~Sun, S.~Jin, J.~Wang, y.~Zhang, X.~Gao, and K.~K. Wong, ``On scheduling for
  massive distributed {MIMO} downlink,'' in \emph{IEEE GLOBECOM'2013}, Atlanta,
  USA, Dec. 2013, pp. 4151--4156.

\bibitem{06095627}
H.~Huh, A.~M. Tulino, and G.~Caire, ``Network {MIMO} with linear zero-forcing
  beamforming: Large system analysis, impact of channel estimation, and
  reduced-complexity scheduling,'' \emph{IEEE Trans. Inf. Theory}, vol.~58,
  no.~5, pp. 2911--2934, May 2012.

\bibitem{m07037316}
B.~Lee, L.~Ngo, and B.~Shim, ``Antenna group selection based user scheduling
  for massive {MIMO} systems,'' in \emph{IEEE GLOBECOM'2014}, Austin, TX, USA,
  Dec 2014, pp. 3302--3307.

\bibitem{m06941317}
G.~Lee and Y.~Sung, ``Asymptotically optimal simple user scheduling for massive
  {MIMO} downlink with two-stage beamforming,'' in \emph{IEEE SPAWC'2014},
  Toronto, Canada, June 2014, pp. 60--64.

\bibitem{06850064}
L.~Sun and M.~R. McKay, ``Tomlinson-harashima precoding for multiuser {MIMO}
  systems with quantized {CSI} feedback and user scheduling,'' \emph{IEEE
  Trans. Signal Process.}, vol.~62, no.~16, pp. 4077--4090, Aug. 2014.

\bibitem{05456039}
E.~Conte, S.~Tomasin, and N.~Benvenuto, ``A simplified greedy algorithm for
  joint scheduling and beamforming in multiuser {MIMO} {OFDM},'' \emph{IEEE
  Commun. Lett.}, vol.~14, no.~5, pp. 381--383, May 2010.

\bibitem{05336871}
R.~H.~Y. Louie, M.~R. McKay, and I.~B. Collings, ``Maximum sum-rate of {MIMO}
  multiuser scheduling with linear receivers,'' \emph{IEEE Trans. Commun.},
  vol.~57, no.~11, pp. 3500--3510, Nov. 2009.

\bibitem{01312486}
D.~Gesbert and M.~S. Alouini, ``How much feedback is multi-user diversity
  really worth?'' in \emph{IEEE ICC'2004}, vol.~1, Paris, France, June 2004,
  pp. 234--238.

\bibitem{01664083}
X.~Shao and J.~Yuan, ``Multiuser scheduling for {MIMO} broadcast and multiple
  access channels with linear precoders and receivers,'' \emph{IEE Proc.
  Commun.}, vol. 153, no.~4, pp. 541--547, August 2006.

\bibitem{04635028}
S.~Lee, J.~Thompson, and J.~Kim, ``Statistical {CSI}-assisted multiuser
  scheduling in {MIMO} broadcast channels,'' \emph{Elec. Lett.}, vol.~44,
  no.~20, pp. 1222--1223, Sept. 2008.

\bibitem{04570215}
Y.~Zhang, C.~Ji, Y.~Liu, W.~Q. Malik, D.~O'brien, and D.~J. Edwards, ``A low
  complexity user scheduling algorithm for uplink multiuser {MIMO} systems,''
  \emph{IEEE Trans. Wireless Commun.}, vol.~7, no.~7, pp. 2486--2491, July
  2008.

\bibitem{04299613}
C.~J. Chen and L.~C. Wang, ``Performance analysis of scheduling in multiuser
  {MIMO} systems with zero-forcing receivers,'' \emph{IEEE J. Sel. Areas
  Commun.}, vol.~25, no.~7, pp. 1435--1445, Sept. 2007.

\bibitem{05374079}
A.~Razi, D.~J. Ryan, I.~B. Collings, and J.~Yuan, ``Sum rates, rate allocation,
  and user scheduling for multi-user {MIMO} vector perturbation precoding,''
  \emph{IEEE Trans. Wireless Commun.}, vol.~9, no.~1, pp. 356--365, Jan. 2010.

\bibitem{05871799}
L.~Jin, X.~Gu, and Z.~Hu, ``Low-complexity scheduling strategy for wireless
  multiuser multiple-input multiple-output downlink system,'' \emph{IET
  Commun.}, vol.~5, no.~7, pp. 990--995, May 2011.

\bibitem{06504540}
E.~Driouch and W.~Ajib, ``Downlink scheduling and resource allocation for
  cognitive radio {MIMO} networks,'' \emph{IEEE Trans. Veh. Technol.}, vol.~62,
  no.~8, pp. 3875--3885, Oct. 2013.

\bibitem{06878495}
W.~Ni, R.~P. Liu, J.~Biswas, X.~Wan, I.~B. Collings, and S.~K. Jha, ``Multiuser
  {MIMO} scheduling for mobile video applications,'' \emph{IEEE Trans. Wireless
  Commun.}, vol.~13, no.~10, pp. 5382--5395, Oct. 2014.

\bibitem{06740115}
C.~Sun, J.~Ge, J.~Li, and B.~Zhu, ``Low complexity user scheduling algorithm
  for energy-efficient multiuser multiple-input multiple-output systems,''
  \emph{IET Commun.}, vol.~8, no.~3, pp. 343--350, Feb. 2014.

\bibitem{06725595}
X.~Yu, X.~Dang, S.~H. Leung, Y.~Liu, and X.~Yin, ``Unified analysis of
  multiuser scheduling for downlink {MIMO} systems with imperfect {CSI},''
  \emph{IEEE Trans. Wireless Commun.}, vol.~13, no.~3, pp. 1344--1355, March
  2014.

\bibitem{05999732}
X.~Yi and E.~K.~S. Au, ``User scheduling for heterogeneous multiuser {MIMO}
  systems: A subspace viewpoint,'' \emph{IEEE Trans. Veh. Technol.}, vol.~60,
  no.~8, pp. 4004--4013, Oct. 2011.

\bibitem{06810624}
N.~Prasad, H.~Zhang, H.~Zhu, and S.~Rangarajan, ``Multi-user {MIMO} scheduling
  in the fourth generation cellular uplink,'' in \emph{IEEE Asilomar'2013},
  Pacific Grove, CA, USA, Nov. 2013, pp. 1855--1859.

\bibitem{04784356}
B.~Song, R.~L. Cruz, and L.~B. Milstein, ``Exploiting multiuser diversity for
  fair scheduling in {MIMO} downlink networks with imperfect channel state
  information,'' \emph{IEEE Trans. Commun.}, vol.~57, no.~2, pp. 470--480, Feb.
  2009.

\bibitem{5452208}
Y.~Polyanskiy, H.~V. Poor, and S.~Verdu, ``Channel coding rate in the finite
  blocklength regime,'' \emph{IEEE Trans. Inf. Theory}, vol.~56, no.~5, pp.
  2307--2359, May 2010.

\bibitem{6127644}
V.~Boussemart, M.~Berioli, F.~Rossetto, and M.~Joham, ``On the achievable rates
  for the return-link of multi-beam satellite systems using successive
  interference cancellation,'' in \emph{Proc. IEEE MILCOM'2011}, Baltimore,
  USA, Nov. 2011, pp. 217--223.

\bibitem{6477555}
A.~Chelli and M.~Alouini, ``Performance of hybrid-{ARQ} with incremental
  redundancy over relay channels,'' in \emph{IEEE GLOBECOM'2012}, Anaheim, CA,
  USA, Dec. 2012, pp. 116--121.

\bibitem{1421925}
J.~G. Andrews, ``Interference cancellation for cellular systems: a contemporary
  overview,'' \emph{IEEE Wireless Commun.}, vol.~12, no.~2, pp. 19--29, April
  2005.

\bibitem{7801976}
B.~Makki, A.~Ide, T.~Svensson, T.~Eriksson, and M.~S. Alouini, ``A genetic
  algorithm-based antenna selection approach for large-but-finite {MIMO}
  networks,'' \emph{IEEE Trans. Veh. Technol.}, vol.~66, no.~7, pp. 6591--6595,
  July 2017.

\bibitem{dominantcdf}
V.~S. Bawa, ``Optimal rules for ordering uncertain prospects,'' \emph{J.
  Financ. Econ.}, vol.~2, no.~1, pp. 95--121, 1975.

\bibitem{5426254}
F.~Yilmaz and M.-S. Alouini, ``Product of the powers of generalized
  {Nakagami-m} variates and performance of cascaded fading channels,'' in
  \emph{IEEE GLOBECOM'2009}, Honolulu, Hawaii, USA, Nov. 2009, pp. 1--8.

\bibitem{FoxHbook}
A.~Erdelyi, W.~Magnus, F.~Oberhettinger, and F.~G. Tricomi, \emph{Higher
  Transcendental Functions}.\hskip 1em plus 0.5em minus 0.4em\relax vol. I,
  McGraw-Hill, New York-Toronto-London, 1953.

\bibitem{1327795}
B.~M. Hochwald, T.~L. Marzetta, and V.~Tarokh, ``Multiple-antenna channel
  hardening and its implications for rate feedback and scheduling,'' \emph{IEEE
  Trans. Inf. Theory}, vol.~50, no.~9, pp. 1893--1909, Sept. 2004.

\bibitem{7134725}
B.~Makki, T.~Svensson, and M.~Zorzi, ``Finite block-length analysis of spectrum
  sharing networks using rate adaptation,'' \emph{IEEE Trans. Commun.},
  vol.~63, no.~8, pp. 2823--2835, Aug. 2015.

\bibitem{4444444444}
T.~M. Cover and J.~A. Thomas, \emph{Elements of Information Theory}.\hskip 1em
  plus 0.5em minus 0.4em\relax New York: Wiley Interscience, 1992.

\bibitem{phdthesisBjornemo}
E.~Bjornemo, ``Energy constrained wireless sensor networks: communication
  principles and sensing aspects,'' {P}h.D. dissertation, Uppsala University,
  Uppsala, Sweden, 2009.

\bibitem{4160747}
S.~Mikami, T.~Takeuchi, H.~Kawaguchi, C.~Ohta, and M.~Yoshimoto, ``An
  efficiency degradation model of power amplifier and the impact against
  transmission power control for wireless sensor networks,'' in \emph{Proc.
  {IEEE} RWS'2007}, Long Beach, CA, USA, Jan. 2007, pp. 447--450.

\bibitem{6725577}
D.~Persson, T.~Eriksson, and E.~G. Larsson, ``Amplifier-aware multiple-input
  single-output capacity,'' \emph{IEEE Trans. Commun.}, vol.~62, no.~3, pp.
  913--919, March 2014.

\bibitem{7104158}
B.~Makki, T.~Svensson, T.~Eriksson, and M.~Nasiri-Kenari, ``On the throughput
  and outage probability of multi-relay networks with imperfect power
  amplifiers,'' \emph{IEEE Trans. Wireless Communications}, vol.~14, no.~9, pp.
  4994--5008, Sept. 2015.

\bibitem{8003368}
M.~Haghifam, M.~Robat~Mili, B.~Makki, M.~Nasiri-Kenari, and T.~Svensson,
  ``Joint sum rate and error probability optimization: Finite blocklength
  analysis,'' \emph{IEEE Wireless Commun. Lett.}, vol.~6, no.~6, pp. 726--729,
  Dec. 2017.

\end{thebibliography}
\vfill
\end{document}